\documentclass[12pt]{iopart}

\usepackage{hyperref}

\usepackage{iopams}  

\usepackage[utf8]{inputenc}
\usepackage{amssymb}
\expandafter\let\csname equation*\endcsname\relax
\expandafter\let\csname endequation*\endcsname\relax

\usepackage{graphicx}
\usepackage{cite}
\usepackage{mathtools} 
\usepackage{amsmath}
\usepackage{amsthm}
\usepackage{mathrsfs}
\usepackage{amsfonts}
\usepackage[font=small,labelfont=bf, width=0.8\linewidth]{caption}

\usepackage{etoolbox}

\makeatletter
\def\@mkboth#1#2{}
\newlength\appendixwidth
\preto\appendix{\addtocontents{toc}{\protect\patchl@section}}
\newcommand{\patchl@section}{%
  \settowidth{\appendixwidth}{\textbf{Appendix }}%
  \addtolength{\appendixwidth}{1.5em}%
  \patchcmd{\l@section}{1.5em}{\appendixwidth}{}{\ddt}%
}
\makeatother

\def\beq{\begin{equation}}
\def\eeq{\end{equation}}
\def\beqa{\begin{eqnarray}}
\def\eeqa{\end{eqnarray}}
\def\bsp#1\esp{\begin{split}#1\end{split}}
\newcommand{\nn}{\nonumber}

\font\cyr=wncyr8
\newcommand{\sha}{{\mbox{\cyr X}}}
\newfont{\scyr}{wncyr10 scaled 550}
\newcommand{\ssha}{\mbox{\bf \scyr X}}
\def\eqn#1{eq.~(\ref{#1})}
\def\eqns#1#2{eqs.~(\ref{#1}) and~(\ref{#2})}
\def\eqnss#1#2{eqs.~(\ref{#1})--(\ref{#2})}
\def\Eqn#1{Eq.~(\ref{#1})}

\newcommand{\cC}{\mathcal{C}}
\def\cL{{\cal L}}
\def\cM{{\cal M}}
\def\cO{{\cal O}}
\def\cR{{\cal R}}
\def\cG{{\cal G}}
\newcommand{\xG}{\mathscr{G}}
\def\Gc{\Gamma_{\rm cusp}}
\def\e{\epsilon}
\def\Sigmat{\tilde{\Sigma}}

\newcommand\sss{\scriptscriptstyle}
\newcommand\as{\alpha_{\sss S}} 
\newcommand\gs{g_{\sss S}}

\newcommand\ws{w^\ast}
\newcommand\zb{\bar{z}}
\newcommand{\Hb}[0]{{\overline{H}}}

\begin{document}
\bibliographystyle{iopart-num}

%%%%XXXX extra command to fix hyperlinks:
%
\newcommand{\eprint}[2][]{\href{https://arxiv.org/abs/#2}{#2}}
%
%%%%XXXX

\begin{flushright}
	SAGEX-22-16 \\
        SLAC-PUB-17670
\end{flushright}

\title[]{}
\title[The Multi-Regge Limit]{The SAGEX Review on Scattering Amplitudes \\
  Chapter 15: The Multi-Regge Limit}
\author{Vittorio Del Duca}

\address{Institute for Theoretical Physics, ETH Z\"urich, 8093 Z\"urich,
  Switzerland\\ Physik-Institut, Universit\"at Z\"urich, 8057 Z\"{u}rich, Switzerland
  \footnote{On leave from INFN, Laboratori Nazionali di Frascati, Italy} }
\ead{delducav@itp.phys.ethz.ch}

\author{Lance J. Dixon}

\address{SLAC National Accelerator Laboratory, Stanford University,
  Stanford, CA 94309, USA}
\ead{lance@slac.stanford.edu}

\vspace{10pt}
\begin{indented}
\item[]March 2022
\end{indented}

\begin{abstract}
We review the Regge and multi-Regge limit of scattering amplitudes
in gauge theory, focusing on QCD and its maximally supersymmetric cousin,
planar ${\cal N}=4$ super-Yang-Mills theory.
We identify the large logarithms that are developed in these limits,
and the progress that has been made in resumming them,
towards next-to-next-to-leading logarithms for BFKL evolution in QCD,
as well as all-orders proposals in planar ${\cal N}=4$ super-Yang-Mills theory
and the perturbative checks of those proposals.  We also cover
the application of single-valued multiple polylogarithms to this important
kinematical limit of particle scattering.
\end{abstract}

%\maketitle

\tableofcontents

\section{Introduction}
\label{sec:intro}

The Regge limit~\cite{Regge:1959mz} of $2\to 2$ scattering amplitudes is defined as the limit in which the squared center-of-mass energy $s$ is much larger than the momentum transfer $t$. In the Regge limit, amplitudes are dominated by the exchange in the $t$ channel of the particle of highest spin.  In non-Abelian gauge theory, that particle is the gluon, or more generally,
the vector boson carrying an $SU(N_c)$ Yang-Mills interaction.
The analysis of the Regge limit in scattering processes in quantum field theories dates back over half a century. It has centered around two concepts: the Reggeization of a particle~\cite{Gribov:1962fw,Gell-Mann:1962xxa,Mandelstam:1965zz,Grisaru:1973vw,Grisaru:1973wbb}, 
understood as the exponentiated $s^{\alpha(t)}$ behavior of the radiative corrections to the $2\to 2$ amplitude when $s\gg |t|$, which is entirely due to the particle exchanged in the $t$ channel, where $\alpha(t)$ is called the Regge trajectory of that particle; and the exchange of a pomeron, i.e.~the behavior of 
cut forward scattering amplitudes under color-singlet exchange in the $t$ channel~\cite{Gribov:1961fr,Chew:1961ev,Cheng:1970xm,Gribov:1970ik}.

In gauge theories, those early studies reached a milestone with the seminal work of Balitsky, Fadin, Kuraev and Lipatov (BFKL), who established
Reggeization of the gluon in $2\to 2$ scattering~\cite{Lipatov:1976zz}; analyzed the behavior of multi-loop multi-leg amplitudes in the multi-Regge limit, in which the produced particles are strongly ordered in rapidity~\cite{Fadin:1975cb,Kuraev:1976ge,Kuraev:1977fs}; and resummed the leading logarithmic (LL) radiative corrections, of ${\cal O}((\as\ln(s/|t|))^n)$, through the BFKL equation~\cite{Kuraev:1977fs,Balitsky:1978ic}.  The BFKL equation describes the behavior of the multi-leg amplitude, squared and integrated over all the allowed final states, which through the optical theorem is equivalent to the $s$-channel cut forward amplitude.
In particular, at $t=0$ the optical theorem relates the square of a multi-leg amplitude with single Reggeized-gluon ladder exchange
to the imaginary part of the $2\to 2$ amplitude with the exchange of a ladder of two Reggeized gluons in a color singlet in the $t$ channel; the latter is
referred to as exchange of the (perturbative) pomeron at $t=0$.

\begin{figure}
  \centerline{\includegraphics[width=0.65\columnwidth]{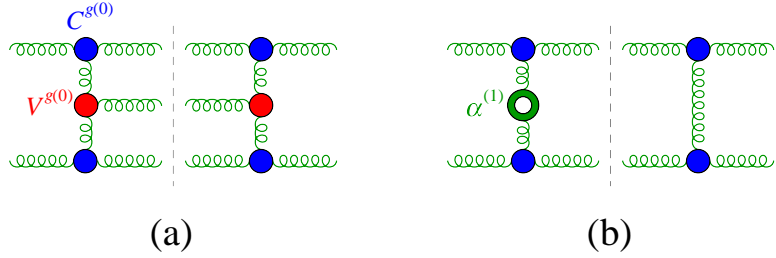} }
  \caption{(a) The red blob along the gluon ladder represents the central--emission vertex within the tree-level five--gluon amplitude.
  (b) The pierced green blob represents the one--loop gluon Regge trajectory within the one--loop four--gluon amplitude. Momentum in the $s$ channel flows horizontally, and in the $t$ channel vertically.}
\label{fig:llbfkl}
\end{figure}
\begin{figure}
  \centerline{\includegraphics[width=0.8\columnwidth]{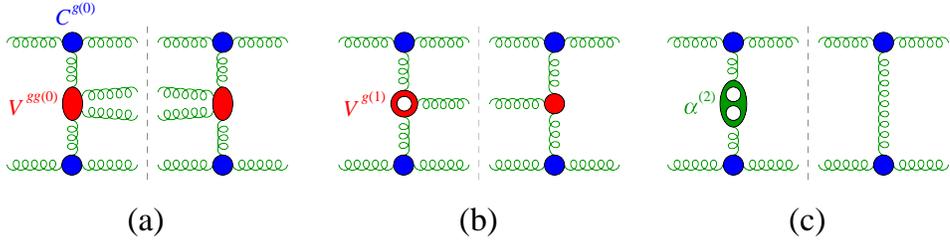} }
  \caption{(a) The red blob along the gluon ladder represents the two-gluon central-emission vertex within the tree six-gluon amplitude.
  (b) The pierced red blob represents the one-loop central-emission vertex within the one-loop five--gluon amplitude.
  (c) The twice pierced green blob represents the two-loop gluon Regge trajectory within the two-loop four--gluon amplitude.}
\label{fig:nllbfkl}
\end{figure}

The BFKL equation is an integral equation with an iterative structure. Its kernel is derived by singling out the emission of a gluon along the gluon ladder, fig.~\ref{fig:llbfkl}(a). The infrared divergences of the kernel, which result from integrating the gluon momentum over its phase space, are regulated by the infrared structure of the one-loop gluon Regge trajectory, fig.~\ref{fig:llbfkl}(b). It is possible to extend the BFKL equation to next-to-leading logarithmic (NLL) accuracy~\cite{Fadin:1998py,Ciafaloni:1998gs,Kotikov:2000pm,Kotikov:2002ab}, i.e.~to resum the radiative corrections of ${\cal O}( \as (\as\ln(s/|t|))^n)$, by considering the radiative corrections to the leading-order kernel. These corrections involve the emission of two gluons, or a $q\bar q$ pair, close in rapidity along the gluon ladder~\cite{Fadin:1989kf,DelDuca:1995ki,Fadin:1996nw,DelDuca:1996nom,DelDuca:1996km}, 
fig.~\ref{fig:nllbfkl}(a), 
and the one-loop corrections to the emission of a gluon along the ladder~\cite{Fadin:1993wh,Fadin:1994fj,Fadin:1996yv,DelDuca:1998cx,Bern:1998sc}, fig.~\ref{fig:nllbfkl}(b).
The infrared divergences of the next-to-leading-order (NLO) kernel, which result from integrating the momenta of the partons emitted along the gluon ladder over their phase space, are regulated by the infrared structure of the two-loop gluon Regge trajectory, fig.~\ref{fig:nllbfkl}(c).

Underpinning the BFKL equation at NLL accuracy is the fact that gluon Reggeization holds at that accuracy~\cite{Fadin:2006bj,Fadin:2015zea}.
Gluon Reggeization breaks down beyond NLL accuracy, because at next-to-next-to-leading logarithmic (NNLL) accuracy three-Reggeized-gluon exchanges appear~\cite{DelDuca:2001gu,DelDuca:2011wkl,DelDuca:2011ae,DelDuca:2013ara,DelDuca:2014cya,Fadin:2016wso,Caron-Huot:2017fxr,Fadin:2017nka,Falcioni:2020lvv,Falcioni:2021buo,Falcioni:2021dgr}. The issue of whether a single-Reggeized-gluon exchange can be isolated and iterated through a BFKL kernel at NNLL accuracy remains to be understood; see sec.~\ref{sec:nnll}.

In the last decade, the study of the multi-Regge limit has deepened after
the realization that it is a powerful kinematic constraint 
for amplitudes in QCD~\cite{Almelid:2017qju,Falcioni:2021buo,Caola:2021izf}
and in the maximally supersymmetric gauge theory, ${\cal N}=4$ super-Yang-Mills theory
(SYM)~\cite{DelDuca:2009au,DelDuca:2010zg,Dixon:2011pw,Dixon:2013eka,Dixon:2014iba,Henn:2016jdu,Caron-Huot:2016owq,Caron-Huot:2019vjl,Caron-Huot:2020vlo},
and that in the Regge limit amplitudes in planar
${\cal N}=4$ SYM~\cite{Dixon:2012yy,Basso:2014pla,DelDuca:2016lad,DelDuca:2019tur}, and amplitudes~\cite{Caron-Huot:2020grv} and cross sections~\cite{DelDuca:2013lma,DelDuca:2017peo} in QCD are endowed with a rich mathematical 
structure.  Although we will not be able to cover them adequately in
this review, effective field theory methods have been brought to bear
on the Regge limit, including the role of Glauber gluons and
quarks~\cite{Rothstein:2016bsq,Moult:2017xpp}; they promise to lead to
further progress on the systematic understanding of this limit in the future.

The multi-Regge limit has been studied extensively in
${\cal N}=4$ SYM, particularly in the limit of a large number of colors,
$N_c\to\infty$, where planar Feynman diagrams dominate.
In this introduction, we provide a review of some of these
developments, prior to going into more detail on many of the topics in
sec.~\ref{sec:planarN4regge}.

In the planar limit, scattering amplitudes all have a definite cyclic
color ordering, with distinct color lines in the fundamental
${\bf N_c}$ representation
of $SU(N_c)$ flowing along each edge.
The color quantum numbers of a Reggeized object exchanged in a given channel,
bounded by two oppositely-oriented edges,
are ${\bf N_c} \otimes {\bf \overline{N}_c}
= {\bf (N_c^2-1)} \oplus {\bf 1}$, but the singlet
contribution is suppressed by a factor of $1/N_c^2$.
Hence the BFKL ladder
studied in planar ${\cal N}=4$ SYM is for the adjoint representation,
whereas QCD BFKL evolution is usually studied at the cross section level
for the singlet channel.
The richness of $n$-gluon scattering amplitudes in planar ${\cal N}=4$
SYM begins at $n=6$, due to an additional dual conformal symmetry
present in the theory~\cite{Drummond:2006rz,Bern:2006ew,
  Alday:2007hr,Bern:2007ct,Drummond:2007aua,Drummond:2007cf,Nguyen:2007ya,
  Bern:2008ap, Drummond:2008aq}.
Because of this symmetry, the four- and five-gluon amplitudes are
completely constrained kinematically to be given by the
Bern-Dixon-Smirnov (BDS) ansatz~\cite{Bern:2005iz},
essentially the exponential of the one-loop amplitude,
because it solves an anomalous dual
conformal Ward identity~\cite{Drummond:2007cf}. 

Starting at $n=6$, the Ward identity allows for non-trivial functions
of the kinematics, which depend on $3n-15$ dual conformal cross ratios.
The first concrete indication that the BDS ansatz had to be modified
at $n=6$ and at two loops came from multi-Regge kinematics (MRK),
where it was shown that the ansatz violates Regge factorization
for both $2\to4$ and $3\to3$ scattering in appropriate
channels~\cite{Bartels:2008ce,Bartels:2009vkz}.
Soon thereafter,
the all-orders factorized structure for $2\to4$ scattering in MRK
was presented for the maximally-helicity-violating (MHV) configuration
in terms of an inverse Fourier-Mellin (FM) transform
of the exponentiated {\it BFKL eigenvalue}
in the adjoint representation,
multiplied by the product of {\it impact factors} for the top
and bottom of the Reggeized ladder~\cite{Lipatov:2010ad,Fadin:2011we}.
The case of $3\to3$ scattering was described in ref.~\cite{Bartels:2010tx};
although closely related to the $2\to4$ case, it is slightly simpler
because the Regge cut contribution in MRK is purely imaginary.
The case of next-to-MHV (NMHV) helicities was analyzed at leading logarithms
in ref.~\cite{Lipatov:2012gk}, and the all-orders factorized structure
was described in refs.~\cite{Basso:2014pla,Dixon:2014iba}.

The full power of integrability in planar ${\cal N}=4$ SYM was
first brought to bear on the six-point MRK limit~\cite{Basso:2014pla}
by performing an intricate analytic continuation from the
pentagon operator product expansion (POPE), or flux tube, representation
of the near-collinear limit~\cite{Basso:2013vsa}.
All-orders predictions were obtained for the adjoint BFKL eigenvalue
and the impact factor, or in other words for all subleading logarithms
at leading power in MRK, for both MHV and NMHV six-point
amplitudes~\cite{Basso:2014pla}.

At each perturbative order, the inverse FM sum
can be computed, and compared to the multi-Regge limit of
amplitudes constructed in general kinematics.
At two loops, the analytic form of the six-point MHV amplitude
was found by explicit computation
of a Wilson loop representation of the
amplitude~\cite{DelDuca:2009au,DelDuca:2010zg}, which was simplified
down to just a few lines using the {\it symbol}
associated with polylogarithmic functions~\cite{Goncharov:2010jf}.
The three- and four-loop MHV and two-, three- and four-loop NMHV
amplitudes were bootstrapped using {\it hexagon functions} with
the correct branch cuts, as well as boundary information from the
near-collinear limit~\cite{Dixon:2011pw,Dixon:2011nj,Caron-Huot:2011dec,
  Dixon:2013eka,Dixon:2014voa,Dixon:2015iva}.
The introduction of constraints on the function space from (extended)
Steinmann relations has made it possible to push as far as seven
loops~\cite{Caron-Huot:2016owq,Caron-Huot:2019vjl,Caron-Huot:2019bsq}.
In some cases the multi-Regge limit has been used to constrain the
bootstrap ansatz; however, the information used is self-consistent,
in the sense that it only requires loop orders in the BFKL eigenvalue
and the impact factor that are already determined by the
amplitude at the previous loop order.
See Chapter 5~\cite{Papathanasiou:2022lan} of the
SAGEX Review~\cite{Travaglini:2022uwo}
for more details about the amplitude bootstrap.

In order to compare the perturbative results to the predictions from
the inverse FM transform, it is helpful to realize that
the six-point results can always be expressed~\cite{Dixon:2012yy}
in terms of real analytic, or single-valued, harmonic polylogarithms
(SVHPLs) for a single complex variable~\cite{BrownSVHPLs}.  At higher points,
multiple-variable SVHPLs appear, which are real analytic functions
on the moduli space of Riemann spheres with marked points or
punctures~\cite{DelDuca:2016lad,Broedel:2016kls}.
Once the inverse FM
transform is known for various building blocks in the FM
representation, they can be combined using a convolution
theorem~\cite{DelDuca:2016lad}.
The inverse FM transform can often be performed
by brute force, by doing it as a truncated series expansion
and matching the result to the series expansion of a 
general linear combination of SVHPLs of the appropriate
weight~\cite{Dixon:2012yy}.  Other algorithms are given in
refs.~\cite{Drummond:2015jea,Broedel:2015nfp}.
Using such methods, the six-point MRK limit predicted by
ref.~\cite{Basso:2014pla} has been verified
through seven loops for both MHV and
NMHV helicity
configurations~\cite{Caron-Huot:2019vjl,ToAppearNMHVSevenLoops}.

Multi-Regge limits of planar ${\cal N}=4$ SYM
amplitudes with more than six external legs
have also received great attention, starting with $2\to5$
scattering at leading logarithmic
accuracy~\cite{Bartels:2011ge,Bartels:2013jna,Bartels:2014jya}.
Besides the same impact factors and BFKL eigenvalue appearing
in the six-point case, a new ${\cal N}=4$
ingredient, the {\it central emission vertex}
(or {\it Lipatov vertex}), first appears for $n=7$ in the so-called
``long'' Regge cut configuration.
The factorized structure beyond leading logarithms was described
and the central emission vertex was obtained at next-to-leading order
in ref.~\cite{DelDuca:2018hrv}.
Based on higher-order perturbative data, and the general structure of
the near-collinear limit, a proposal for the all-orders
form of the central emission vertex was presented,
and its perturbative predictions
were checked at the symbol level through four loops
for the MHV configuration~\cite{DelDuca:2019tur},
relying on the amplitudes bootstrapped in
refs.~\cite{Drummond:2014ffa,Dixon:2016nkn,Drummond:2018caf}.
Recently the proposal was checked through four loops
at full function level for both MHV and NMHV seven-point
amplitudes~\cite{Dixon:2021nzr}, making use of the zeta-valued
constants fixed in ref.~\cite{Dixon:2020cnr}.

Beyond seven points, it is possible that no new ingredients are
required for amplitudes in the long Regge cut configuration. 
This is the case at two loops, at least at the level
of the symbol of the MHV $n$-point amplitude, which has been
computed in generic kinematics~\cite{Caron-Huot:2011zgw},
and studied in MRK~\cite{Prygarin:2011gd,Bargheer:2015djt,DelDuca:2019tur}.
However, as discussed further in the conclusions,
for amplitudes in other cut configurations there still may be more to learn from double
and higher discontinuities at two loops~\cite{DelDuca:2018raq} and beyond~\cite{Bartels:2020twc}.

At strong coupling, scattering amplitudes in planar ${\cal N}=4$
SYM are given in terms of the area of a minimal surface in five-dimensional
anti-de Sitter space that is bounded by a polygon composed
of light-like edges~\cite{Alday:2007hr}.
The minimal area problem is integrable and can be solved using
a thermodynamic Bethe ansatz or
$Y$-system~\cite{Alday:2009yn,Alday:2009dv,Alday:2010vh}.
These systems have been solved in multi-Regge
limits~\cite{Bartels:2010ej,Bartels:2012gq,Bartels:2013dja,
  Bartels:2014ppa,Bartels:2014mka,Sprenger:2016jtx,Abl:2021hhb},
shedding light on the strong-coupling behavior, which at six-points
must be consistent with the strong coupling limit of the
all-orders results~\cite{Basso:2014pla}.

The remainder of this review is organized as follows.
In sec.~\ref{sec:qcdregge}, we consider the multi-Regge limit of QCD amplitudes, the BFKL equation, its solution and the function space which describes it, at LL and at NLL accuracy. At the end of the section, we comment briefly on ongoing work beyond NLL accuracy.
In sec.~\ref{sec:planarN4regge}, we analyze the multi-Regge limit of amplitudes with six and seven points in planar ${\cal N}=4$ SYM.
We describe the conformal cross ratios, the symbol alphabet, the function space, and the all-orders formulae which are supposed to hold for amplitudes at six and more points. In sec.~\ref{sec:concl}, we draw our conclusions and briefly discuss the integrability picture of amplitudes in ${\cal N}=4$ SYM in the large $N_c$ limit.

%%%%%%%%%%%%%%%%%%%%%%%%%%%%%%%%%%%%%%%%%%%%%%%%

\section{The multi-Regge limit of QCD amplitudes}
\label{sec:qcdregge}

\begin{figure}
  \centerline{ \includegraphics[width=0.3\columnwidth]{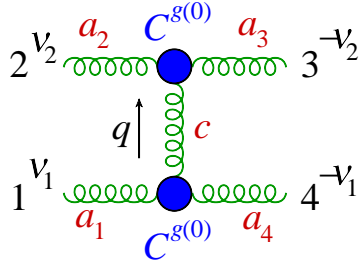} }
  \caption{Amplitude for gluon-gluon scattering in the Regge limit. Momenta and helicities are labeled in black, color in red. 
    The blue blobs represent the impact factors.
    The helicity labelling stresses that, at tree level and
    at leading power in $t/s$,
    helicity is conserved along the horizontal $s$-channel direction.}
\label{fig:tree4g}
\end{figure}

In the Regge limit, $s\gg |t|$, $2\to 2$ scattering amplitudes in QCD are dominated by gluon exchange in the $t$ channel. Contributions which do not feature gluon exchange in the $t$ channel
are power suppressed in $t/s$. At tree level we can write the $2\to 2$ amplitudes in a factorized way. For example, 
the tree amplitude for gluon-gluon scattering $g_1\, g_2\to g_3\,g_4$ in the helicity basis\footnote{We take all the momenta as outgoing, so the helicity labels for incoming partons are the negative of their physical helicities.}
may be written as \cite{Lipatov:1976zz,Kuraev:1976ge}, 
\begin{equation}
\cM^{(0)}_{4g} = 
\left[\gs (F^{a_3})_{a_2c}\, C^{g(0)}(p_2^{\nu_2}, p_3^{\nu_3})) \right]
{s\over t} \left[\gs (F^{a_4})_{a_1c}\, C^{g(0)}(p_1^{\nu_1}, p_4^{\nu_4}) \right]\, ,\label{elas}
\end{equation}
with momenta $p_2$ in the $+$ light-cone direction
and $p_1$ in the $-$ light-cone direction, as shown in fig.~\ref{fig:tree4g}.
We use light-cone coordinates adapted to the incoming beam directions,
$p^{\pm}= p_0\pm p_z $, and complexified transverse momenta
$p_\perp = p_x+ip_y$, $p^*_\perp = p_x-ip_y$. Hence a momentum vector
$p_i^\mu = (p_i^+,p_i^-,p_{i\perp})$ has Lorentz norm
$p_i^2 = p_i^+ p_i^- - |p_{i\perp}|^2$, and
$2p_i\cdot p_j = p_i^+ p_j^- + p_i^- p_j^+
- p_{i\perp} p_{j\perp}^* - p_{i\perp}^* p_{j\perp}$.
In general, we denote external momenta by $p_i$ (occasionally $k_i$)
and reserve $q_i$ for $t$-channel momentum exchanges between factorized
emissions.  In the present four-point case, we define
$s=(p_1+p_2)^2$, $q = p_2 + p_3$, $t=q^2  \simeq -|q_\perp|^2$,
The superscripts $\nu_i$ label the helicities.
The adjoint generators of the gauge group are the structure constants,
$(F^c)_{ab} = i\sqrt{2} f^{acb}$.
It is apparent from the color coefficient $(F^{a_3})_{a_2c} (F^{a_4})_{a_1c}$
in \eqn{elas} that only the antisymmetric octet ${\bf 8}_a$
is exchanged in the $t$ channel.

Because the four-gluon amplitude is a MHV amplitude, \eqn{elas} describes
$\binom{4}{2} = 6$ helicity configurations. However, at tree level and at
leading power in $t/s$, helicity is conserved along the $s$-channel
direction shown in fig.~\ref{fig:tree4g},
or in our all-outgoing helicity convention,
\beq
C^{g}(p_2^{\nu_2}, p_3^{\nu_3})\ \propto\ \delta^{\nu_2,-\nu_3} .
\label{Chelcons}
\eeq
Thus in \eqn{elas} four helicity configurations are leading,
two for each tree-level impact factor, $g^*\, g \rightarrow g$,
with $g^*$ an off-shell gluon~\cite{DelDuca:1995zy},
\begin{equation}
C^{g(0)}(p_2^\ominus, p_3^\oplus) = 1\,, \qquad C^{g(0)}(p_1^\ominus, p_4^\oplus) = {p_{4\perp}^* \over p_{4\perp}}\, ,\label{centrc}
\end{equation} 
with complex transverse coordinates $p_{\perp} = p^x + i p^y$.\footnote{
The apparent asymmetry under the flip $p_1\leftrightarrow p_2$,
$p_3\leftrightarrow p_4$ is just an external-state phase convention.}
At this order, the impact factors are just overall phases, and they 
transform under parity into their complex conjugates,
\begin{equation}
[C^{g}(p_i^\nu, p_j^{\nu'})]^* = C^{g}(p_i^{-\nu}, p_j^{-\nu'})\, . 
\end{equation} 
The helicity-flip impact factor $C^{g(0)}(p_i^\oplus, p_j^\oplus)$ and its parity conjugate $C^{g(0)}(p_i^\ominus, p_j^\ominus)$ 
are power suppressed in $t/s$. However, helicity flip terms along the $s$-channel direction, and thus helicity--violating impact factors, do occur at one
loop~\cite{Fadin:1993wh,Fadin:1992zt,DelDuca:1998kx}.

The tree amplitudes for quark-gluon or quark-quark scattering have the same form as \eqn{elas}, up to replacing one or both
gluon impact factors $C^{g(0)}$ in \eqn{centrc}
with quark impact factors $C^{q(0)}$,
and the color factors $(F^c)_{ab}$ in the adjoint representation
with the color factors $T^c_{ij}$ in the fundamental representation of $SU(3)$,
which we normalize as $\Tr(T^aT^b)= T_F \delta^{ab}$, with $T_F=1$.
So in the Regge limit, the $2\to 2$ scattering amplitudes factorize into gluon or quark impact factors
and a gluon propagator in the $t$ channel, and are uniquely determined by them.

The loop corrections to an amplitude feature poles and branch cuts, which are dictated by the analytic
structure and constrained by the symmetries of the amplitude. 
In the Regge limit $s\simeq -u \gg -t$, $2\to 2$ scattering amplitudes are symmetric under $s\leftrightarrow u$ crossing.
Thus we may consider amplitude combinations whose kinematic and color coefficients have a definite signature under $s\leftrightarrow u$ crossing,
\begin{equation}
\cM_4^{(\pm)}(s,t) = \frac{ \cM_4(s,t) \pm \cM_4(u,t) }{2}\,,
\label{eq:sucross}
\end{equation}
with $u=-s-t\simeq -s$, such that $\cM_4^{(-)}(s,t)$ ($\cM_4^{(+)}(s,t)$) has kinematic and color coefficients which are both odd (even) 
under $s\leftrightarrow u$ crossing.
Furthermore, higher-order contributions to $g g \rightarrow g g$ scattering
in general involve additional color structures, as dictated by the decomposition of the product ${\bf 8}_a \otimes {\bf 8}_a$
into irreducible representations,
\beq
  {\bf 8}_a \otimes {\bf 8}_a \, = \, \{ {\bf 1} \oplus {\bf 8}_s \oplus {\bf 27} \} 
  \oplus [ {\bf 8}_a \oplus {\bf 10} \oplus \overline{{\bf 10}} ]\ \, ,
\label{8*8}
\eeq
where in curly (square) brackets are the representations which are even (odd) under $s\leftrightarrow u$ crossing.

\subsection{The Regge limit at leading logarithmic accuracy}
\label{sec:reggell}

When loop corrections to the tree amplitude (\ref{elas}) are considered, 
it is found that at leading logarithmic (LL) accuracy in $\ln(s/|t|)$, the four-gluon amplitude is given to all orders
in $\as$ by~\cite{Lipatov:1976zz,Kuraev:1976ge}
\begin{equation}
\left. \cM_{4g}\right|_{LL} =  \left[\gs (F^{a_3})_{a_2c}\, C^{g(0)}(p_2^{\nu_2}, p_3^{\nu_3})) \right] {s\over t}
\left({s\over \tau}\right)^{\alpha(t)}\, \left[\gs (F^{a_4})_{a_1c}\, C^{g(0)}(p_1^{\nu_1}, p_4^{\nu_4}) \right]\,,
\label{sud}
\end{equation}
where $\tau > 0$ is a Regge factorization scale, which is of order of $t$, although the precise definition of $\tau$ 
is immaterial for four-point amplitudes or to LL accuracy,
where one can suitably fix $\tau = - t$. In \eqn{sud}, $\alpha(t)$ is called
the Regge trajectory. It is given by an integral over the loop
transverse momentum,
\begin{equation}
\alpha(t) = \as\, C_A\, t \int {d^2k_{\perp}\over (2\pi)^2}\, {1\over k_{\perp}^2
(q-k)_{\perp}^2}\, ,\label{allv}
\end{equation}
with $\as = \gs^2/(4\pi)$ and $C_A=N_c$ the number of colors.
Regulating the integral in $d=4-2\epsilon$ dimensions, one obtains
\begin{equation}
\alpha(t) =\frac{N_c \as}{4\pi}\alpha^{(1)}(t)\,, \qquad {\rm with} \qquad 
\alpha^{(1)}(t) =  {\gamma_K^{(1)}\over 4\epsilon} 
\left(\mu^2\over -t\right)^{\epsilon} \kappa_{\Gamma}\, ,\label{alph}
\end{equation}
with
\begin{equation}
\kappa_\Gamma = (4\pi)^\epsilon\, {\Gamma(1+\epsilon)\,
\Gamma^2(1-\epsilon)\over \Gamma(1-2\epsilon)}\, ,\label{cgam}
\end{equation}
and where $\gamma_K^{(1)}$ is the one-loop coefficient of the cusp anomalous dimension~\cite{Korchemsky:1985xj,Moch:2004pa}, 
\beq
\label{hatgammaK}
  \gamma_K (\as)  = \sum_{\ell=1}^\infty \gamma_K^{(\ell)} \left( \frac{N_c \as}{4\pi} \right)^\ell\,, \quad {\rm with} \quad
  \gamma_K^{(1)} = 8 \,.
\eeq
Although no renormalization occurs at LL accuracy, in \eqn{alph} the renormalization scale $\mu$ appears and provides a scaling dimension.
Its presence is understood henceforth.

The prominent features of eq.~(\ref{sud}) are that at LL accuracy the amplitude (\ref{sud}) is still real,
the antisymmetric octet ${\bf 8}_a$ is still the only color representation exchanged in the $t$ channel,
\begin{equation}
\left. \cM_{4g}\right|_{LL} = \left. \cM^{(-)[8_a]}_{4g}\right|_{LL}\,, \qquad\qquad \left. \cM^{(+)}_{4g}\right|_{LL} = 0\,,
\end{equation}
and the one-loop result (\ref{alph}) exponentiates.
The exponentiation of $\ln(s/|t|)$ in the one-loop result, which effectively dresses the gluon propagator as
\begin{equation}
\frac1{t} \to \frac1{t} \left({s\over \tau}\right)^{\alpha(t)}\,,
\label{eq:reggeize}
\end{equation}
is called gluon Reggeization, and we say that in the Regge limit the four-gluon amplitude~(\ref{sud}) features the exchange in the
$t$ channel of one Reggeized gluon.

Because of \eqn{sud}, factorization holds at LL accuracy just like at
tree level, i.e.~the 
amplitudes for quark-gluon or quark-quark scattering at LL accuracy
have the same form as \eqn{sud}, up to replacing one or both
color and impact factors for gluons with the ones for quarks.

\subsection{The Multi-Regge limit}
\label{sec:multiregge}

The Regge limit of the $2\to 2$ amplitudes in \eqn{elas} is characterized by strong orderings in the light-cone momenta of the two final-state gluons,
\beq
p_3^+\gg p_4^+\,, \qquad p_3^-\ll p_4^-\,,
\label{eq:strordpm}
\eeq
where the second strong ordering is equivalent to the first because of the mass-shell conditions $p_i^+p_i^- = |p_{i\perp}|^2$, with $i=3, 4$,
and of transverse momentum conservation, $p_{3\perp} + p_{4\perp} = 0$.
Since for a light-like momentum, $p^\pm = |p_\perp| e^{\pm y}$, where $y$ is the rapidity,
\eqn{eq:strordpm} is equivalent to a strong ordering of the rapidities of the final-state gluons.

Next we consider $2\to 3$ amplitudes with momenta $p_1\, p_2\to p_3\,p_4\, p_5$.
Here the Regge limit is realized by the two kinematic limits,
\beq
p_3^+\gg p_4^+\simeq p_5^+ \quad {\rm or} \quad p_3^+\simeq p_4^+\gg p_5^+\,, \qquad {\rm with } \qquad
|p_{3\perp}| \simeq |p_{4\perp}| \simeq |p_{5\perp}|\,.
\label{eq:nmrk12}
\eeq
The two kinematics of \eqn{eq:nmrk12}
are termed next-to-multi-Regge kinematics (NMRK).
They have an overlap in the kinematic region characterized by a strong ordering in the light-cone momenta of all three final-state gluons,
\beq
p_3^+\gg p_4^+\gg p_5^+\,, \qquad {\rm with } \qquad |p_{3\perp}| \simeq |p_{4\perp}| \simeq |p_{5\perp}|\,,
\label{eq:mrk}
\eeq
which is called multi-Regge kinematics (MRK).
In MRK, the tree amplitude for five-gluon scattering $g_1\, g_2\to g_3\,g_4\, g_5$ takes the factorized ladder form,
\begin{eqnarray}
\cM^{(0) }_{5g}  &=& 
s \left[\gs (F^{a_3})_{a_2c_1}\, C^{g(0)}(p_2^{\nu_2}, p_3^{\nu_3})) \right]\, 
{1\over t_1} \label{three}\\ &\times& \left[\gs (F^{a_4})_{c_1c_2}\, 
V^{g(0)}(q_1,p^{\nu_4}_4,q_2) \right]\, {1\over t_2}\, 
\left[\gs (F^{a_5})_{a_1c_2}\, C^{g(0)}(p_1^{\nu_1}, p_5^{\nu_5}) \right]\, .\nn
\end{eqnarray}
with $q_1 = p_2+p_3$, $q_2= q_1+p_4$, and $t_i= q_i^2\simeq - q_{i\perp} q_{i\perp}^\ast$, with $i = 1, 2$, where the impact factors are given in \eqn{centrc}.
The emission of a gluon along the gluon ladder is governed by the
{\it central-emission vertex} (CEV)~\cite{Lipatov:1976zz,Lipatov:1991nf},
\beq
V^{g(0)}(q_1,p^\oplus_4,q_2) = \frac{q_{1\perp}^\ast q_{2\perp}}{p_{4\perp}}\,.
\label{eq:lipv}
\eeq
Note that while \eqn{three} displays soft divergences in the limit that gluon $p_4 \to 0$, collinear divergences are screened by the MRK,
\eqn{eq:mrk}, which prevents the invariant mass of any two partons from becoming arbitrarily small.

\subsection{The BFKL equation at LL accuracy}
\label{sec:bfkl}

The ladder form of \eqn{three} can be iterated to provide
the tree amplitude for $n$-gluon scattering in MRK, 
\beq
p_3^+\gg p_4^+\gg \ldots \gg p_n^+\,, 
\label{eq:mrkn}
\eeq
where a requirement on the transverse momenta to be all of the same size is understood, by adding $n-5$ central-emission vertices
along the ladder of amplitude (\ref{three}). The ensuing tree-level $n$-gluon amplitude, with $n-4$ central-emission vertices and $n-3$ gluon propagators,
is uplifted to all orders in $\as$, at LL accuracy in $\ln(s/|t|)$, by dressing each of the gluon propagators as in \eqn{eq:reggeize}.
Just like the four-gluon amplitude (\ref{sud}) in the Regge limit, the $n$-gluon amplitude in MRK at LL accuracy is characterized by
the exchange of one Reggeized gluon, which is termed the Reggeon.

The central-emission vertex (\ref{eq:lipv}), fig.~\ref{fig:llbfkl}(a), and the gluon Reggeization (\ref{eq:reggeize}),
fig.~\ref{fig:llbfkl}(b), constitute the building blocks of an iterative structure, which is captured by the BFKL equation~\cite{Fadin:1975cb,Kuraev:1976ge,Kuraev:1977fs,Balitsky:1978ic},
which sums the terms of ${\cal O}(\as^n \ln^n(s/|t|))$ and
describes the evolution of a gluon ladder in transverse momentum and in rapidity.
In the BFKL equation, real emissions as well as virtual ones are included.
In order to match the LL accuracy of the virtual corrections (\ref{sud}),
amplitudes with five or more gluons are taken in MRK (\ref{eq:mrk}),
as in \eqn{three}.  
The MRK rationale is that each gluon emitted along the ladder requires a factor of $\as$, and the integral over its rapidity yields a factor of $\ln(s/|t|)$, so that each emitted gluon contributes a factor of ${\cal O}(\as \ln(s/|t|))$.

We can display how the BFKL equation works by considering gluon-gluon scattering. 
In the Regge limit, at leading order in $\as$, i.e.~${\cal O}(\as^2)$, the partonic cross section for gluon-gluon scattering $g_1\, g_2\to g_3\,g_4$ 
is~\cite{DelDuca:1995hf} 
\begin{equation}
{d\hat\sigma_{gg}^{(0)}\over d^2 p_{3\perp} d^2 p_{4\perp}}\ =\
\biggl[{N_c\as\over |p_{3\perp}|^2}\biggr] \,
{1\over2}\delta^{(2)}(p_{3\perp}+p_{4\perp}) \,
\biggl[{N_c\as\over |p_{4\perp}|^2}\biggr] \,,
\label{cross0}
\end{equation}
which is obtained by squaring amplitude (\ref{elas}) and integrating it over the phase space 
of the final-state gluons 3 and 4.
The terms in square brackets are related to the square of the impact factors (\ref{centrc}), which is just 1, multiplied by an overall factor.
The real corrections in $\as$, i.e.~${\cal O}(\as^3)$, are obtained by squaring the five-gluon amplitude (\ref{three}), whose momenta we re-label
as $p_1 p_2\to p_3 k_1 p_4$, and integrating it over the phase space of the final-state gluons~\cite{DelDuca:1994ng},
\begin{equation}
{d\hat\sigma_{gg}^{(1r)}\over d^2 p_{3\perp} d^2 p_{4\perp}}\ =\
\biggl[{N_c\as\over |p_{3\perp}|^2}\biggr] \,
\frac{N_c\as}{\pi^2} \int \frac{ d^2 k_{1\perp} d y_{k_1}}{k_{1\perp}^2}
{1\over2}\delta^{(2)}(p_{3\perp}+k_{1\perp}+p_{4\perp}) \,
\biggl[{N_c\as\over |p_{4\perp}|^2}\biggr] \ ,
\label{cross1r}
\end{equation}
where the superscript (1r) on the left-hand-side stands for real radiation of the first loop order. In \eqn{cross1r},
$y_{k_1}$ is integrated over the range $\Delta y= y_3-y_4 =\ln(s_{12}/|p_{3\perp}| |p_{4\perp}|)$. The integral over $k_{1\perp}$ yields a logarithmic
soft singularity, which is regulated by including the virtual corrections, \eqns{sud}{allv}. The finite remainder is a term of ${\cal O}(\as \Delta y)$.

The subsequent orders in $\as$ each yield an integral over transverse momentum with a weight $\frac{N_c\as}{\pi^2} \int \frac{{\rm d}^2 k_{i\perp} }{k_{i\perp}^2}$, and an integral over rapidity bounded as in \eqn{eq:mrkn}, which for the ${\cal O}(\as^{n+2})$ corrections yield a factor of $\frac{(\Delta y)^n}{n!}$.
Including all the orders of ${\cal O}(\as \Delta y)$, the gluon-gluon initiated cross section in the Regge limit can be written as~\cite{Mueller:1986ey,DelDuca:1993mn,Stirling:1994he}
\begin{equation}
{d\hat\sigma_{gg}\over d^2 p_{3\perp} d^2 p_{4\perp}}\ =\
\biggl[{N_c\as\over |p_{3\perp}|^2}\biggr] \,
f(q_{1\perp},q_{2\perp},\Delta y) \,
\biggl[{N_c\as\over |p_{4\perp}|^2}\biggr] \ ,
\label{cross}
\end{equation}
where, as in \eqn{three}, $q_{1\perp} = p_{3\perp}$ and $q_{2\perp} = - p_{4\perp}$. 
$f(q_{1\perp},q_{2\perp},\Delta y)$ is the solution of the BFKL equation for
evolution in rapidity,
\beq
\frac{\partial}{\partial \Delta y} f(q_{1\perp},q_{2\perp},\Delta y)
\ =\ \left( {\cal K}\star f\right) (q_{1\perp}, q_{2\perp},\Delta y) \,,
\label{eq:BFKLequation}
\eeq
which can be given an explicit iterative form by writing it
as~\cite{Schmidt:2001yq}
\beqa
f(q_{1\perp},q_{2\perp},\Delta y) &=& {1\over2}\delta^{(2)}(p_{3\perp}+p_{4\perp}) \nn\\
&+& \Delta y\, {\cal K}\left[ {1\over2}\delta^{(2)}(p_{3\perp}+p_{4\perp}) \right] \nn\\
&+& \frac{(\Delta y)^2}2 {\cal K}\left[ {\cal K}\left[ {1\over2}\delta^{(2)}(p_{3\perp}+p_{4\perp}) \right] \right]\nn\\
&+& \frac{(\Delta y)^2}{3!} {\cal K}\left[ {\cal K}\left[ {\cal K}\left[ {1\over2}\delta^{(2)}(p_{3\perp}+p_{4\perp}) \right] \right] \right]\nn\\
&+& \ldots \,.
\label{eq:bfklmc}
\eeqa
The integral operator ${\cal K}$ is a convolution,
\beq
{\cal K}\left[ f(q_{1\perp},q_{2\perp}) \right] = \left( {\cal K}\star f\right) (q_{1\perp}, q_{2\perp}) =
\int d^2 k_{\perp} K(q_{1\perp}, k_\perp) f(k_\perp, q_{2\perp})\,,
\label{eq:intop}
\eeq
with
\beqa
\hspace{-2cm} && \left( {\cal K}\star f\right) (q_{1\perp}, q_{2\perp}) \nn \\ \hspace{-2cm} && \qquad = 
\frac{N_c\as}{\pi^2} \int d^2 k_{\perp} \frac1{|q_{1\perp}-k_{\perp}|^2} 
\left( f(k_\perp, q_{2\perp}) - {|q_{1\perp}|^2\over |k_{\perp}|^2 + |q_{1\perp}-k_{\perp}|^2} f(q_{1\perp}, q_{2\perp}) \right)
%K(q_{1\perp}, q_{2\perp}) = \frac{N_c\as}{\pi^2} \frac1{|q_{1\perp}-q_{2\perp}|^2}
%\left( 1 - {|q_{1\perp}|^2\over |q_{2\perp}|^2 + |q_{1\perp}-q_{2\perp}|^2}\, \delta^{(2)}(q_{1\perp}-q_{2\perp}) \right)
,\label{kern}
\eeqa
where the first term corresponds to the emission of a gluon along the ladder, \eqns{eq:lipv}{cross1r}, and the second term to the virtual 
corrections, \eqns{sud}{allv} (after partial fractioning
and a change of integration variable).

The kernel $K$ is obtained by generalizing \eqn{three}
to $n$-gluon scattering in MRK and by Reggeizing, 
like in \eqn{eq:reggeize}, each of the $(n-3)$ ensuing gluon propagators~\cite{Fadin:1975cb,Kuraev:1976ge,Kuraev:1977fs} in order to 
obtain the $n$-gluon amplitude at LL accuracy. This is then squared
(the square of the CEV (\ref{eq:lipv}) will yield the first term of \eqn{kern})
and integrated over the phase space of the $(n-2)$ outgoing gluons.
The rapidities are integrated over, while the $(n-2)$ integrals over
transverse momentum can be written as a recursive relation through
the integral operator (\ref{eq:intop})~\cite{DelDuca:1995hf}. 
Using the fact that the square of the CEV (\ref{eq:lipv}) is regular as
$q_{2\perp}\to\infty$ and vanishes as $q_{2\perp}\to 0$,
it is possible to show~\cite{DelDuca:1995hf} that  \eqn{kern} 
and thus the solution (\ref{eq:bfklmc}) of the BFKL equation are regular in the
ultraviolet and in the infrared regimes, respectively.

The solution (\ref{eq:bfklmc}) of the BFKL equation is amenable to a
Monte Carlo implementation of the gluon 
ladder~\cite{Schmidt:1996fg,Orr:1997im,Andersen:2011hs}. The resummed form
of the solution~\cite{Kuraev:1977fs,Balitsky:1978ic}
is obtained by transforming it to moment space, 
\begin{equation}
f(q_{1\perp},q_{2\perp},\Delta y) \ =\ \int {d\omega\over 2\pi i}\, 
e^{\omega\Delta y}\, 
f_{\omega}(q_{1\perp},q_{2\perp})\label{moment}
\end{equation}
such that we can write the BFKL equation as
\begin{equation}
\omega\, f_{\omega}(q_{1\perp},q_{2\perp})\, =
{1\over 2}\,\delta^{(2)}(q_{1\perp}-q_{2\perp})\, +
\left( {\cal K}\star f_{\omega}\right) (q_{1\perp}, q_{2\perp})
\, ,\label{bfklb}
\end{equation}
with the kernel $K$ as in \eqn{kern}. The BFKL equation is solved by finding a set of eigenfunctions $\Phi_{\nu n}(q)$
of the integral operator ${\cal K}$,
\beq
\left( {\cal K}\star \Phi_{\nu n}\right) (q_{\perp}) = \omega_{\nu n} \Phi_{\nu n} (q_{\perp})\,,
\label{eq:bfklnun}
\eeq
where $\nu$ is a real number, $n$ is an integer, and $\omega_{\nu n}$ is the BFKL eigenvalue~\cite{Balitsky:1978ic}.
In a conformally-invariant theory,
the eigenfunctions $\Phi_{\nu n}(q)$ are fixed by conformal symmetry~\cite{Lipatov:1985uk}.
They coincide with the eigenfunctions $\varphi_{\nu n}(q)$ of QCD at LL accuracy, 
\beq
\label{loeigenf}
\Phi_{\nu n}(q)\equiv\varphi_{\nu n}(q) = {1\over 2\pi}\, (q^2)^{-1/2+i\nu}\, e^{i n \theta}\,,
\eeq
where $\theta$ is the azimuthal angle of $q$, and they satisfy the completeness relation,
\beq
\sum_{n=-\infty}^{+\infty} \int_{-\infty}^{+\infty} d\nu\, \Phi_{\nu n}(q)\, \Phi_{\nu  n}^\ast({q^\prime}) 
= {1\over 2}\,\delta^{(2)}(q-q^\prime) = \delta\left(q^2-{q^\prime}^2\right)\, \delta(\theta-\theta^\prime) \,.
\label{complete}
\eeq

In terms of the eigenfunctions (\ref{loeigenf}) and the eigenvalue in \eqn{eq:bfklnun}, the solution to the BFKL equation (\ref{bfklb})
can be written as
\begin{equation}
f_\omega(q_1,q_2)\, = \sum_{n=-\infty}^{+\infty} \int_{-\infty}^{+\infty} d\nu\, {1\over \omega -\omega_{\nu n}}\,
\Phi_{\nu n}(q_1)\, \Phi_{\nu n}^\ast(q_2)\,.
\label{solb}
\end{equation}
In fact, we can apply the integral operator to \eqn{solb},
\beq
({\cal K}\star f_\omega)(q_1,q_2) = \sum_{n=-\infty}^{+\infty} \int_{-\infty}^{+\infty} d\nu\, {1\over \omega -\omega_{\nu n}}\,
({\cal K}\star\Phi_{\nu n})(q_1)\, \Phi_{\nu n}^\ast(q_2) \,.
\label{bfklid}
\eeq
Then using \eqn{eq:bfklnun} and the completeness relation (\ref{complete}) in \eqn{bfklid}, the BFKL equation (\ref{bfklb}) is identically satisfied.

Using eq.~(\ref{moment}) on the solution (\ref{solb}) of the BFKL equation, we can write it as
\beq
f(q_{1\perp},q_{2\perp},\Delta y) = \sum_{n=-\infty}^{\infty} \int_{-\infty}^{\infty} d\nu\, \Phi_{\nu n}(q_1)\,
 \Phi^\ast_{\nu n}(q_2)\, e^{\Delta y\,\omega_{\nu n}} \,,\label{sola}
\eeq
which, using the eigenfunctions (\ref{loeigenf}), becomes
\beq
f(q_{1\perp},q_{2\perp},\Delta y) = 
{1\over (2\pi)^2 \sqrt{ |q_{1\perp}|^2 |q_{2\perp}|^2 } }
\sum_{n=-\infty}^{\infty} e^{in\phi}\, \int_{-\infty}^{\infty} d\nu\, 
e^{\eta\, \chi_{\nu n}}\, \left(|q_{1\perp}|^2\over |q_{2\perp}|^2
\right)^{i\nu}\, ,\label{solc}
\eeq
where $\phi$ is the angle between $q_{1\perp}$ and $q_{2\perp}$.
The exponent in \eqn{solc} is given by
$\eta\, \chi_{\nu n} = \Delta y\,\omega_{\nu n}$, with
\beq
\eta = \frac{N_c \as}{\pi} \Delta y \,,
\label{eq:etadef}
\eeq
where the BFKL eigenvalue is
\beq
\omega_{\nu n} = \frac{N_c\as}{\pi}\,\chi_{\nu n}\,,
\label{eq:llbfkl}
\eeq
In order to find the explicit form of the eigenvalue $\chi_{\nu n}$ in \eqn{eq:llbfkl}, we replace the solution (\ref{solb}) 
with the eigenfunctions (\ref{loeigenf}) into the homogeneous part of the BFKL equation (\ref{bfklb}), and we obtain
\beqa
 \chi_{\nu n} &=& 2\,{\rm Re}\int_0^1 dx\, 
{x^{{|n|-1\over 2}+i\nu} \over 1-x} - 2 \int_0^1 dx\, {1\over 1-x} 
\nn\\ &-& \int_0^1 dx\, {1\over x}\, +\, \int_0^1 dx\, {1\over x
\sqrt{1+4x^2}}\, +\, \int_0^1 dx\, {1\over \sqrt{x^2+4}}\, ,\label{spec}
\eeqa
with
\beq
x =\, \left\{ \begin{array}{ll} 
q_2^2/q_1^2 & \mbox{for $q_2^2 < q_1^2$}\, ,\\ 
q_1^2/q_2^2 & \mbox{for $q_2^2 > q_1^2$}\, .\end{array} 
\right. \label{xvar} 
\eeq
The last three terms in eq.~(\ref{spec}) cancel out, and the eigenvalue becomes 
\beq
 \chi_{\nu n} = -2\gamma_E
-\psi\left(\frac{1}{2}+\frac{|n|}{2}+i\nu\right)
-\psi\left(\frac{1}{2}+\frac{|n|}{2}-i\nu\right)\,,
\label{eq:chinun}
\eeq
where $\gamma_E = - \psi(1)$ is the Euler-Mascheroni constant and 
\beq
{d\ln{\Gamma(y)}\over dy}\, =\, \psi(y)\, =\, \int_0^1 dx {x^{y-1}-1\over x-1}
- \gamma_E \, \label{gam}
\eeq
is the logarithmic derivative of the $\Gamma$ function.

Note that the kernel $K$ (\ref{kern}) is real and symmetric, 
so the integral operator ${\cal K}$ (\ref{eq:intop}) is hermitian and its eigenvalue (\ref{eq:llbfkl}) is real.
In addition, in \eqn{eq:llbfkl} there are no beta function terms, in 
accordance with the lack of collinear or ultraviolet divergences in the BFKL kernel. Accordingly, the BFKL eigenvalue
at LL accuracy is the same in QCD and in ${\cal N}=4$ SYM. Finally, in the BFKL eigenvalue (\ref{eq:llbfkl}) there are 
only leading $N_c$ terms.

The solution (\ref{solc}) of the BFKL equation at LL accuracy can be expanded into a power series in $\eta$,
\beq
f(q_{1\bot},q_{2\bot},\Delta y) 
= \frac{1}{2}\delta^{(2)}(q_{1\bot}- q_{2\bot})
+\frac{1}{2\pi\, \sqrt{|q_{1\perp}|^2 |q_{2\perp}|^2} }
\,\sum_{k=1}^\infty\eta^k \,f_k(w,\ws)\,,
\label{eqn:greensvhpl}
\eeq
where $w$ is a complex variable,
\beq
w = \frac{p_{3\perp} }{p_{4\perp} }\,,
\label{eq:w}
\eeq
such that
\beq
|w|^2 = \frac{|p_{3\bot}|^2}{|p_{4\bot}|^2} = \frac{|q_{1\bot}|^2}{|q_{2\bot}|^2}
{\rm~~~~and~~~~}
\left(\frac{w}{\ws}\right)^{1/2} 
= e^{-i\phi_{jj}} = -e^{i\phi}\,,
\label{eq:variab}
\eeq
where $\phi_{jj} = \pi - \phi$ is the angle between $p_{3\perp}$ and $p_{4\perp}$.
In \eqn{eqn:greensvhpl}, the coefficients $f_k$ are given by the FM transform,
\beq
f_k(w,\ws) = {\cal F}[ \chi_{\nu n}^k] = 
\frac{1}{k!}
\sum_{n=-\infty}^{+\infty}(-1)^n\,\left(\frac{w}{\ws}\right)^{n/2}
\int_{-\infty}^{+\infty}\frac{d\nu}{2\pi}\,|w|^{2i\nu}\,\chi_{\nu n}^k\,.
\label{eq:fmc}
\eeq
The coefficients $f_k$ are real-analytic functions of $w$, that is,
they have a unique, well-defined value for every ratio
of the magnitudes of the two jet transverse momenta and angle between
them. Furthermore, \eqn{eq:fmc} is invariant under $n\leftrightarrow-n$ and
$\nu\leftrightarrow-\nu$, which implies that the $f_k$ are invariant under
conjugation and inversion of $w$,
\beq\label{eq:fk_symmetry}
f_k(w,\ws) = f_k(\ws,w) = f_k(1/w,1/\ws)\,,
\eeq
i.e.~the coefficients $f_k$ are eigenfunctions under
the action of the $\mathbb{Z}_2\times\mathbb{Z}_2$ symmetry generated by
\beq\label{eq:Z2xZ2}
(w,\ws) \leftrightarrow (\ws,w) {\rm~~~~and~~~~}
(w,\ws) \leftrightarrow (1/w,1/\ws)\,.
\eeq
A special point in the $(w,\ws)$ plane
is at $w = \ws = -1$, which corresponds to the Born kinematics,
where the two jets are back-to-back,
with equal and opposite transverse momentum,
$|p_{3\bot}|^2 = |p_{4\bot}|^2$ and $\phi_{jj} = \pi$.
Another special point is the origin, $w = \ws = 0$, when one jet
has much smaller transverse momentum than the other jet.
The point at infinity is related to the origin by the inversion symmetry,
while $w = \ws = -1$ is a fixed point of the
$\mathbb{Z}_2\times\mathbb{Z}_2$ symmetry~\eqref{eq:Z2xZ2}.

In analogy with the multi-Regge limit of the
six-point MHV and NMHV amplitudes in ${\cal N}=4$
SYM theory~\cite{Pennington:2012zj},
a generating function can be introduced such as to write the coefficients $f_k$ as~\cite{DelDuca:2013lma} 
\beq
f_k(w,\ws) = \frac{|w|}{|1+w|^2}\,F_k(w,\ws)\,,
\label{fkww}
\eeq
where the pure transcendental functions $F_k$ are given in terms of SVHPLs. 
For example, the first few loop orders of the functions $F_k$ are
\beqa
F_1(w,\ws) &=& 1\,, \nn\\
F_2(w,\ws) &=& - \cL_1 - \frac{1}{2} \cL_0 \,, \nn\\
F_3(w,\ws) &=& \cL_{1,1} + \frac{1}{2} ( \cL_{0,1} + \cL_{1,0} ) + \frac{1}{6} \cL_{0,0} \,.
\label{fcoeffcL}
\eeqa
In order to make contact with the SVHPLs $\cL_{\vec\omega}(z,\bar{z})$ defined in sec.~\ref{sec:SixgluonMRK}, \eqnss{eq:Lwt1}{eq:Lwt2},
we note that the poles of the SVHPLs are at $z=0$ and 1, not $w=0$ and $-1$, so we will need to identify $(z,\bar{z})=(-w,-\ws)$;
thus in \eqn{fcoeffcL} it is understood that $\cL_{\vec\omega}\equiv \cL_{\vec\omega}(-w,-\ws)$.
Finally, using the $\mathbb{Z}_2\times\mathbb{Z}_2$ symmetry~\eqref{eq:Z2xZ2}, projections of the SVHPLs onto eigenstates
under conjugation as well as under inversion can be defined~\cite{Dixon:2012yy}.

Using the all-orders expression for
the perturbative expansion of the BFKL solution (\ref{eqn:greensvhpl}) at LL accuracy,
we can immediately write down the explicit
expression for the gluon-gluon cross section \eqref{cross}
in the Regge limit to any loop order, in LL approximation. In particular, 
we can obtain explicit analytic expressions for the dijet cross section in the Regge limit at
LL accuracy that are inclusive in the transverse
momentum and exclusive in the azimuthal angle, or vice-versa, or inclusive in both.
Accordingly, analytic expressions for the azimuthal-angle distribution and for the transverse-momentum distribution
were obtained~\cite{DelDuca:2013lma}, as well for the case where both
the transverse momenta (above a threshold $E_\bot$)
and the azimuthal angle are integrated over,
the so-called Mueller-Navelet dijet cross section~\cite{Mueller:1986ey},
\beq
\hat\sigma_{gg} = \frac{\pi (N_c\as)^2}{2 E_\bot^2}
\, \sum_{k=0}^\infty f_{0,k}\, \eta^k\,,
\label{eq:mnexp}
\eeq
for which the coefficients $f_{0,k}$ were computed analytically
through the $13^{\rm th}$ order in terms of multiple zeta values~\cite{DelDuca:2013lma}.
As an example, we reproduce here the coefficient of the $13^{\rm th}$ order,
\beqa
\hspace{-1cm} f_{0,13} &=&
\frac{4513}{1890}\,\zeta_{5,3}\,\zeta_5+\frac{27248}{23625}\,\zeta_{5,3,3}\,\zeta_2-\frac{97003}{235200}\,\zeta_{5,5,3}+\frac{13411}{75600}\,\zeta_{7,3}\,\zeta_3 \nn\\ \hspace{-1cm}
&+& \frac{7997743}{12700800}\,\zeta_{7,3,3}-\frac{187318}{14175}\,\zeta_4\,\zeta_3^3-\frac{125056}{4725}\,\zeta_2\,\zeta_5\,\zeta_3^2-\frac{17411413}{302400}\,\zeta_7\,\zeta_3^2 \nn\\ \hspace{-1cm}
&-& \frac{5724191}{100800}\,\zeta_5^2\,\zeta_3-\frac{1874972477}{2376000}\,\zeta_{10}\,\zeta_3-\frac{2418071698069}{2235340800}\,\zeta_{13} \nn\\
\hspace{-1cm} &-& \frac{2379684877}{6048000}\,\zeta_{11}\,\zeta_2-\frac{297666465053}{523908000}\,\zeta_6\,\zeta_7
-\frac{1770762319}{2494800}\,\zeta_5\,\zeta_8 \nn\\ \hspace{-1cm} &-& \frac{229717224973}{628689600}\,\zeta_4\,\zeta_9\,.
\eeqa

\subsection{The Regge limit at NLL accuracy}
\label{sec:reggenll}

At next-to-leading-logarithmic (NLL) accuracy, 
taking into account the $s\leftrightarrow u$ crossing symmetry (\ref{eq:sucross}), 
the exchange of one Reggeized gluon of \eqn{sud}
generalizes to~\cite{Fadin:1993wh}
\beqa
\hspace{-2cm} \lefteqn{ \cM^{(-)[8_a]}_{4g} } \label{sudall}\\
\hspace{-2cm} &=& \frac1{2}  
\left[\gs (F^{a_3})_{a_2c}\, C^{g}(p_2^{\nu_2}, p_3^{\nu_3}) \right] {s\over t} 
\left[ \left({s\over \tau}\right)^{\alpha(t)} + \left({-s\over \tau}\right)^{\alpha(t)} \right]
\, \left[\gs (F^{a_4})_{a_1c}\, C^{g}(p_1^{\nu_1}, p_4^{\nu_4}) \right] \,, \nn
\eeqa
where the color and kinematic parts of the amplitude are each odd under $s\leftrightarrow u$ crossing,
and where we expand in $\as$ the gluon Regge trajectory,
\begin{equation}
\alpha(t) = \frac{N_c\as}{4\pi} \alpha^{(1)}(t) + 
\left(\frac{N_c\as}{4\pi}\right)^2 \alpha^{(2)}(t) + {\cal O}(\as^3)\,
,\label{alphb}
\end{equation}
with $\alpha^{(1)}(t)$ given in \eqn{alph}, and where the (unrenormalized) two-loop coefficient,
in the conventional dimensional regularization (CDR)/'t-Hooft-Veltman (HV) schemes,
is~\cite{Fadin:1995xg,Fadin:1996tb,Fadin:1995km,Blumlein:1998ib,DelDuca:2001gu}
\beq
\alpha^{(2)}(t) = \kappa_{\Gamma}^2 \left(\mu^2\over -t\right)^{2\epsilon} 
\left( \frac{\beta_0}{\epsilon^2} + \frac{\gamma_K^{(2)}}{8\epsilon} 
+ \frac{\gamma_\wedge^{(2)}}2 + \zeta_2 \beta_0
%{404\over 27} - 2\zeta_3 - {56\over 27}\, \frac{N_f}{N_c} 
\right) + {\cal O}(\epsilon)\,,
\label{eq:2looptraj}
\eeq
with $N_f$ the number of light quark flavors,
$\beta_0$ the one-loop coefficient of the beta function,
\beq
\beta_0 = \frac{11}3 - \frac 23\frac{N_f}{N_c}\,,
\label{eq:b0k2}
\eeq
$\gamma_\wedge^{(2)}$ the two-loop coefficient of the ``wedge" anomalous dimension~\cite{Erdogan:2011yc,Falcioni:2019nxk}
for Wilson lines in the adjoint representation, which starts at two loops,
\beq
\gamma_\wedge (\as)  = \sum_{\ell=2}^\infty \gamma_\wedge^{(\ell)} \left( \frac{N_c \as}{4\pi} \right)^\ell , \quad {\rm with} \quad
\gamma_\wedge^{(2)} = {808\over 27} - 4\zeta_3 - {112\over 27}\, \frac{N_f}{N_c} - 2\zeta_2 \beta_0 \,,
\label{eq:wedge}
\eeq
and $\gamma_K^{(2)}$ the two-loop coefficient~\cite{Korchemsky:1987wg,Caron-Huot:2015bja} of the cusp anomalous dimension (\ref{hatgammaK}),
\beq
\gamma_K^{(2)} = 8 \left( \frac{64}9  + \frac{\delta_R}3 - 2\zeta_2 \right) - \frac{80}9 \frac{N_f}{N_c}\,,
\label{eq:k2}
\eeq
where
\begin{equation}
\delta_R = \left\{ \begin{array}{ll} 1 & \mbox{HV/CDR},\\
0 & \mbox{DR/FDH}. \end{array} \right. \label{cp}
\end{equation}
$\delta_R$ is a regularization parameter, which labels the computation as done in CDR/HV schemes for $\delta_R =1$, or
in the dimensional reduction (DR)/ four dimensional helicity (FDH) schemes for $\delta_R =0$.

In \eqn{sudall}, the helicity-conserving impact factor is expanded in $\as$ as
\begin{equation}
C^{g}(p_i^{\nu_i}, p_j^{-\nu_i};\tau) = 
C^{g (0)}(p_i^{\nu_i}, p_j^{-\nu_i})\left(1 + \frac{N_c\as}{4\pi} c^{g (1)}(t;\tau) + {\cal O}(\as^2) \right)\, ,
\label{fullv}
\end{equation}
where the one-loop coefficient $c^{g (1)}$ is real and independent of the helicity configuration.
Its unrenormalized version 
is~\cite{Fadin:1993wh,Fadin:1992zt,Fadin:1993qb,DelDuca:1998kx,Bern:1998sc,DelDuca:2017pmn}
\beqa
\hspace{-2cm} c^{g (1)}(t,\tau) &=&  \kappa_\Gamma \left({\mu^2\over -t}\right)^{\epsilon} \bigg[
- \frac{\gamma_K^{(1)} }{4\epsilon^2} + \frac{\gamma_g^{(1)} }{\epsilon} + \frac{\beta_0}{2\epsilon} 
+ \frac{\gamma_K^{(1)} }{8\epsilon} \ln\left(\frac{\tau}{-t}\right)  
- \frac{\gamma_K^{(2)} }{16} + 2\zeta_2   \nn\\
\hspace{-2cm} && \qquad\qquad\quad 
- \frac12 \left( \frac{\gamma_\wedge^{(2)}}2 + \zeta_2 \beta_0 \right)
%\left(\zeta_3 - {202\over 27} + {28\over 27}\, \frac{N_f}{N_c} \right) 
\epsilon \bigg] + {\cal O}(\epsilon^2) \,,
\label{impactcorrect}
\eeqa
where $\gamma_g^{(1)}$ is the one-loop coefficient of the gluon collinear anomalous dimension,
\beq
  \gamma_g (\as)  = \sum_{\ell=1}^\infty \gamma_g^{(\ell)} \left( \frac{N_c \as}{4 \pi} \right)^\ell\,, \qquad 
  {\rm with} \qquad \gamma_g^{(1)} = - \beta_0\,.
  \label{eq:collad}
\eeq
\Eqn{impactcorrect} is valid in the CDR/HV~\cite{Fadin:1993wh,Fadin:1992zt,DelDuca:1998kx,Bern:1998sc} and 
DR/FDH~\cite{DelDuca:1998kx,Bern:1998sc} schemes through ${\cal O}(\epsilon^0)$, and in the HV scheme
through ${\cal O}(\epsilon)$. 
Expressions to all orders in $\epsilon$ are known in the HV scheme~\cite{Fadin:1993wh,Fadin:1992zt,Bern:1998sc}.
Note that the two-loop trajectory (\ref{eq:2looptraj}) and the one-loop impact factor (\ref{impactcorrect}) are expressed in terms 
of anomalous dimensions which are characteristic of infrared 
factorization in the Regge limit~\cite{DelDuca:2011wkl,DelDuca:2011ae,DelDuca:2013ara,DelDuca:2014cya}.
However, the connection~\cite{DelDuca:2017pmn} between the ${\cal O}(\epsilon^0)$ term of the two-loop
trajectory (\ref{eq:2looptraj}) and the ${\cal O}(\epsilon)$ term of the one-loop impact factor (\ref{impactcorrect})
is as yet unexplained.

\begin{figure}
  \centerline{\includegraphics[width=0.5\columnwidth]{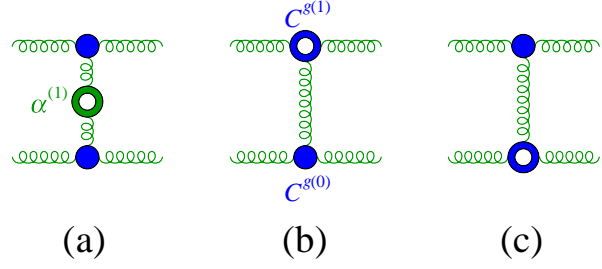} }
  \caption{One-loop factorization of the four-gluon amplitude.
  (a) The one-loop gluon Regge trajectory is represented by the pierced green blob.
  (b) The one-loop gluon impact factor is represented by the pierced blue blob.}
\label{fig:oneloopregge}
\end{figure}
In addition, we may define a signature-symmetric logarithm,
\beq
L = \frac12 \left[ \ln\left({s\over \tau}\right) + \ln\left({-s\over \tau}\right) \right] = \ln\left({s\over \tau}\right) - i\frac{\pi}2\,,
\label{eq:symmlog}
\eeq
and write the four-gluon amplitude (\ref{sudall}) as a double expansion in $\as$ and in $L$, 
\beq
 \cM^{(-)[8_a]}_{4g} =  \cM^{(0)}_{4g}
\left( 1 + \sum_{\ell=1}^\infty \left(\frac{N_c\as}{4\pi}\right)^\ell  \sum_{i=0}^\ell M^{(-,\ell,i)[8_a]}_{4g} L^i \right) \, ,
\label{elasexpand}
\eeq
where $i=\ell$ yields the coefficients at LL accuracy, $i=\ell-1$ the coefficients at NLL accuracy,
and in general $i=\ell-k$ the coefficients at $\mathrm{N^kLL}$ accuracy.

Beyond LL accuracy, in the gluon ladder the exchange of two or more Reggeized gluons may appear.
Furthermore, all the color representations (\ref{8*8}) exchanged in the $t$ channel may contribute.
A similar expansion to \eqn{elasexpand} can be given for the amplitudes $\cM^{(+)}_{4g}$
whose kinematic and color parts are both even under $s\leftrightarrow u$ crossing.
Using the logarithm (\ref{eq:symmlog}), one can show that the coefficients $M^{(\mp,\ell,i)}_{4g}$
of the odd (even) amplitudes are real (imaginary)~\cite{Caron-Huot:2017fxr}, and that the odd (even) amplitudes
display gluon ladders with the $t$-channel exchange of an odd (even) number of Reggeized gluons.

\begin{figure}
  \centerline{\includegraphics[width=0.7\columnwidth]{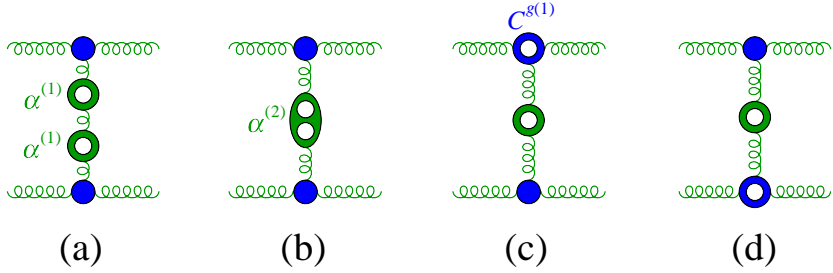} }
  \caption{Two-loop factorization of the four-gluon amplitude at NLL accuracy.
  (a) The one-loop gluon Regge trajectory squared.
  (b) The two-loop gluon Regge trajectory is represented by the twice pierced green blob.
  (c, d) The product of the one-loop trajectory times the one-loop impact factor.}
\label{fig:twoloopregge}
\end{figure}

However, at NLL accuracy, the real part of the amplitude is entirely given by the antisymmetric octet ${\bf 8}_a$
through \eqn{sudall},
\begin{equation}
 {\rm Re} \left[ {\cal M}_{4g} \right]_{NLL} = {\rm Re}\left[ {\cal M}^{(-)[8_a]}_{4g} \right] \,, 
 \label{eq:reamp4gnll}
\end{equation}
which, once \eqn{sudall} is expanded at one and two loops reads,
\beqa
\hspace{-2cm}
{\rm Re}\left[ {\cal M}^{(1)}_{4g} \right]_{NLL} &=& \alpha^{(1)}(t) \ln\left(\frac{s}{\tau}\right) + 2 c^{g (1)}(t,\tau) \,, \label{eq:1loopnll}\\
\hspace{-2cm} {\rm Re}\left[ {\cal M}^{(2)}_{4g} \right]_{NLL} &=&
\frac1{2} \left( \alpha^{(1)}(t) \right)^2 \ln^2\left(\frac{s}{\tau}\right) 
+ \left( \alpha^{(2)}(t) + 2 c^{g (1)}(t,\tau) \alpha^{(1)}(t) \right) \ln\left(\frac{s}{\tau}\right)\,. \label{eq:2loopnll}
\eeqa
\Eqn{eq:1loopnll} is the one-loop factorization of the gluon-gluon amplitude. The single-logarithmic term is the one-loop gluon Regge trajectory, 
fig.~\ref{fig:oneloopregge}(a), which is LL accurate. The non-logarithmic terms are the one-loop impact factors, fig.~\ref{fig:oneloopregge}(b, c), 
which are NLL accurate. \Eqn{eq:2loopnll} is the two-loop factorization of the gluon-gluon amplitude at NLL accuracy. The double-logarithmic term 
is the one-loop trajectory squared, fig.~\ref{fig:twoloopregge}(a), which is LL accurate. The single-logarithmic terms are the two-loop gluon Regge 
trajectory, fig.~\ref{fig:twoloopregge}(b), and the product of the one-loop trajectory times the one-loop impact factors, figs.~\ref{fig:twoloopregge}(c, d).
They are NLL accurate.

Beyond two loops, no more coefficients occur at NLL accuracy,
i.e.~the gluon-gluon scattering amplitude is uniquely
determined by \eqn{sudall}, in terms of the two-loop Regge trajectory 
$\alpha^{(2)}(t)$
and the one-loop impact factor $c^{g(1)}$.
Accordingly, gluon Reggeization is extended to NLL accuracy~\cite{Fadin:2006bj,Fadin:2015zea}.
In addition, because of \eqn{sudall} factorization still holds, so
the amplitudes for quark-gluon or quark-quark scattering 
have the same form as \eqn{sudall}, up to replacing one or both
color and impact factors for gluons with the ones for quarks.

\subsection{The BFKL kernel at NLL accuracy}
\label{sec:kernnll}

In order to extend the BFKL equation beyond the LL accuracy, the kernel of the integral operator (\ref{eq:intop}) is expanded in the strong coupling as
\begin{equation}
K(q_1,q_2) = 4\, \overline{\alpha}_\mu \sum_{\ell=0}^\infty\overline{\alpha}_\mu^\ell\, K^{(\ell)}(q_1,q_2)\,,
\end{equation}
where
\beq
\overline{\alpha}_\mu = \frac{N_c\, \as(\mu^2)}{4\pi}
\label{eq:rescalalpha}
\eeq
is the rescaled renormalized strong coupling constant evaluated at an arbitrary scale $\mu^2$. 
$K^{(0)}$ is the leading-order BFKL kernel~(\ref{kern})~\cite{Fadin:1975cb,Kuraev:1976ge,Kuraev:1977fs,Balitsky:1978ic}, 
which leads to the resummation of the terms of ${\cal O}\left( (\overline{\alpha}_\mu \ln(s/\tau))^n\right)$, i.e., terms at LL accuracy, and the NLO kernel $K^{(1)}$~\cite{Fadin:1998py,Ciafaloni:1998gs} resums the terms at NLL accuracy,
i.e.~of ${\cal O}\left(\overline{\alpha}_\mu (\overline{\alpha}_\mu \ln(s/\tau))^n\right)$, and so forth.

At NLL accuracy, the kernel of the BFKL equation is given by the radiative corrections to the CEV, i.e.~the emission of two gluons or of a
quark-antiquark pair along the 
gluon ladder~\cite{Fadin:1989kf,DelDuca:1995ki,Fadin:1996nw,DelDuca:1996nom,DelDuca:1996km}, 
fig.~\ref{fig:nllbfkl}(a), 
and the one-loop corrections to the CEV~\cite{Fadin:1993wh,Fadin:1994fj,Fadin:1996yv,DelDuca:1998cx,Bern:1998sc}, 
fig.~\ref{fig:nllbfkl}(b).
The infrared divergences of the radiative corrections to the CEV
cancel the divergences of the two-loop Regge trajectory, fig.~\ref{fig:nllbfkl}(c), making the solution of the BFKL equation at NLL accuracy
infrared finite.

In analogy with the tree-level four-gluon amplitude (\ref{elas}), which is upgraded by \eqn{sudall} to an all-orders
expression which is valid at NLL accuracy, we may lift the tree-level five-gluon amplitude in MRK~(\ref{three})
to an all-orders expression at NLL accuracy through the equation,
\begin{eqnarray}
\hspace{-1cm} \cM^{(-,-)[8_a]}_{5g}  &=& \frac1{4}
s \left[\gs (F^{a_3})_{a_2c_1}\, C^{g}(p_2^{\nu_2}, p_3^{\nu_3})) \right]\, 
{1\over t_1} \left[ \left({s_{34}\over \tau}\right)^{\alpha(t_1)} + \left({-s_{34}\over \tau}\right)^{\alpha(t_1)} \right]
\nonumber\\ \hspace{-1cm} &&\times \left[\gs (F^{a_4})_{c_1c_2}\, 
V^{g}(q_1,p^{\nu_4}_4,q_2) \right]\, {1\over t_2}
\left[ \left({s_{45}\over \tau}\right)^{\alpha(t_2)} + \left({-s_{45}\over \tau}\right)^{\alpha(t_2)} \right] \nn\\
\hspace{-1cm} &&\times
\left[\gs (F^{a_5})_{a_1c_2}\, C^{g}(p_1^{\nu_1}, p_5^{\nu_5}) \right]\,, \label{threeall}
\end{eqnarray}
where the $(-,-)$ label on the left-hand side specifies that the amplitude has color and kinematic  coefficients 
which are both odd under both the crossings $p_2\leftrightarrow p_3$ and $p_1\leftrightarrow p_5$.
In \eqn{threeall}, the impact factors are expanded in $\as$ as in \eqn{fullv}, while the CEV is expanded as
\beq
	V^g(q_1,p_4^{\nu_4},q_2;\tau)=V^{g(0)}(q_1,p_4^{\nu_4},q_2)
	\left(1 +\frac{\as}{4\pi}v^{g(1)}(t_1,|p_{4\perp}|^2,t_2;\tau)+\mathcal{O}\left(\as^2\right) \right)\,, \label{eq:allcev}
\eeq
with $V^{g(0)}$ as in \eqn{eq:lipv}. The one-loop corrections 
$v^{g(1)}$ are provided in
refs.~\cite{Fadin:1993wh,Fadin:1994fj,Fadin:1996yv,DelDuca:1998cx,Bern:1998sc}.
They are independent of the regularization scheme choice~\cite{DelDuca:1998cx}.
As outlined above, they contribute to the virtual corrections to the BFKL kernel at NLL accuracy.

The emission of two gluons or of a quark-antiquark pair along the 
gluon ladder requires considering the tree amplitude for six-gluon scattering
$g_1\, g_2\to g_3\,g_4\, g_5\, g_6$ in the NMRK
in which the gluons are strongly ordered in rapidity
except for two gluons (or for
a quark-antiquark pair emitted along the gluon ladder),
\beq
p_3^+\gg p_4^+\simeq p_5^+ \gg p_6^+\,, \qquad {\rm with } \qquad
|p_{3\perp}| \simeq |p_{4\perp}| \simeq |p_{5\perp}| \simeq |p_{6\perp}|\,.
\label{eq:nmrk45}
\eeq
In the NMRK (\ref{eq:nmrk45}) the tree six-gluon amplitude factorizes 
as
\beqa
\hspace{-2cm} {\cal M}^{(0)}_{6g} &=& s\, \sum_{\sigma \in S_2}
\left[\gs (F^{a_3})_{a_2c_1}\, C^{g(0)}(p_2^{\nu_2}, p_3^{\nu_3})) \right]\, 
{1\over t_1} \label{treesix}\\ \hspace{-2cm} 
&\times& \left[\gs^2 (F^{a_{\sigma_4}}F^{a_{\sigma_5}})_{c_1c_3}\, 
V^{gg(0)}(q_1, p^{\nu_{\sigma_4}}_{\sigma_4}, p^{\nu_{\sigma_5}}_{\sigma_5}, q_3) \right]\, {1\over t_3}\, 
\left[\gs (F^{a_6})_{a_1c_3}\, C^{g(0)}(p_1^{\nu_1}, p_6^{\nu_6}) \right] \,, \nn
\eeqa
where the sum is over the permutations of the labels 4 and 5.
The CEV for the emission of two gluons
$V^{gg(0)}(q_1, p^{\nu_4}_4, p^{\nu_5}_5, q_3)$ (or of a quark-antiquark 
pair)~\cite{Fadin:1989kf,DelDuca:1995ki,Fadin:1996nw,DelDuca:1996km}, 
contributes the real corrections to the BFKL kernel at NLL accuracy.
The NMRK rationale is that when the two gluons or the quark-antiquark pair 
are integrated over their common rapidity,
they yield a factor of ${\cal O}( \as^2 \ln(s/\tau) )$, thus
contributing to NLL accuracy.

\subsection{The BFKL equation at NLL accuracy}
\label{sec:bfklnll}

The BFKL eigenvalue and eigenfunctions also admit an expansion in the strong coupling,
\begin{equation}\label{eq:expansion}
\omega_{\nu n} = 4\,\overline{\alpha}_\mu \sum_{\ell=0}^\infty\overline{\alpha}_\mu^\ell\, \omega_{\nu n}^{(\ell)}\,, \qquad
\varphi_{\nu n}(q) = \sum_{\ell=0}^\infty\overline{\alpha}_\mu^\ell\, \varphi_{\nu n}^{(\ell)}(q)\,,
\end{equation}
where $\overline{\alpha}_\mu$ is given in \eqn{eq:rescalalpha},
$\omega_{\nu n}^{(0)} = \chi_{\nu n}$ is given in \eqn{eq:chinun} and $\varphi_{\nu n}^{(0)}(q)$ in \eqn{loeigenf}.
The NLO corrections, $\omega_{\nu n}^{(1)}$, to the BFKL eigenvalue were computed for $n=0$~\cite{Fadin:1998py} and 
later for arbitrary $n$~\cite{Kotikov:2000pm,Kotikov:2002ab}, albeit in the approximation that the NLO eigenfunctions are identical to the LO eigenfunctions given in eq.~\eqref{loeigenf}, 
\begin{equation}
({\cal K}^{NLO}\star\varphi_{\nu n})(q) \equiv
 4\, \overline{\alpha}_{\sss S}(q^2) \left( \chi_{\nu n} + \overline{\alpha}_{\sss S}(q^2) \delta_{\nu n} \right)\, \varphi^{(0)}_{\nu n}(q) +\mathcal{O}(\overline{\alpha}_{\sss S}^3(q^2))\,.
\label{homognlo}
\end{equation}
In this approximation, the NLO corrections to the eigenvalue $\delta_{\nu n}$ in QCD are given by~\cite{Fadin:1998py,Kotikov:2000pm,Kotikov:2002ab},
\beq
\delta_{\nu n} = 6\zeta_3 + \frac{\gamma_K^{(2)} }8\, \chi_{\nu n} + \delta_{\nu n}^{(a)} + \delta_{\nu n}^{(b)} 
+ \delta_{\nu n}^{(c)} - \frac{1}{2}\beta_0\, \chi^2_{\nu n} + \frac{i}{2}\beta_0\, \partial_\nu \chi_{\nu n} \,, \label{nloeigenv}
\eeq
with $\beta_0$ the one-loop coefficient of the beta function in \eqn{eq:b0k2} and $\gamma_K^{(2)}$ the two-loop coefficient of the cusp anomalous dimension in \eqn{eq:k2}.

We split the more complicated contributions into three pieces,
\beqa
\hspace{-2.5cm} \delta_{\nu n}^{(a)} &=& \partial_\nu^2 \chi_{\nu n} \,, \nn\\
\hspace{-2.5cm} \delta_{\nu n}^{(b)} &=& -2\Phi(n,\gamma) - 2\Phi(n,1-\gamma) \,, \nn\\
\hspace{-2.5cm} \delta_{\nu n}^{(c)} &=&
- \frac{\Gamma(\gamma)\Gamma(1-\gamma)}{2i\nu}\left[ \psi \left( \gamma \right) - \psi \left( 1-\gamma \right) \right] \label{nloeigenv3} \\  
\hspace{-2.5cm}
  &\times& \bigg[ \delta_{n0}\left( 3+ \left( 1 + \frac{N_f}{N^3_c} \right)\frac{2 + 3\gamma (1-\gamma)}{(3-2\gamma)(1+2\gamma)} \right) 
  - \delta_{|n|2}\left(\left( 1 + \frac{N_f}{N^3_c} \right)\frac{\gamma(1-\gamma)}{2(3-2\gamma)(1+2\gamma)} \right) \bigg],\nonumber
\eeqa
where we used the shorthand $\gamma = 1/2 + i\nu$, with $\Phi(n,\gamma)$ defined as,
\begin{align}
\Phi(n,\gamma)	&= \sum^{\infty}_{k=0} \frac{(-1)^{k+1}}{k + \gamma + |n|/2}\bigg\{ \psi'(k + |n|+1)-\psi'(k+1) \nonumber\\
& + (-1)^{k+1}[\beta'(k + |n|+1)+\beta'(k+1)] -\frac{1}{k + \gamma + |n|/2}[\psi(k+|n|+1)-\psi(k+1)]\bigg\} \,,
\end{align}
with
\begin{equation}
\beta'(z) = \frac{1}{4}\left[ \psi'\left( \frac{1+z}{2}\right) -\psi'\left( \frac{z}{2}\right) \right] \,.
\end{equation}

For ${\cal N}=4$ SYM the BFKL eigenvalue is given by~\cite{Kotikov:2000pm}
the first four terms of \eqn{nloeigenv}, 
\begin{equation}\label{eq:delta_N=4}
 \delta_{\nu n}^{{\cal N}=4} = 6\zeta_3 + \frac{\gamma_K^{(2)\,{\cal N}=4} }8 \chi_{\nu n} 
+ \partial_\nu^2 \chi_{\nu n} -2\Phi(n,\gamma) - 2\Phi(n,1-\gamma)\,,
\end{equation}
with $\gamma_K^{(2) \, {\cal N}=4}$ the two-loop cusp anomalous dimension in ${\cal N}=4$ SYM,
\begin{equation}\label{eq:cusp_N=4}
\gamma_K^{(2)\,{\cal N}=4} = -16\zeta_2 \,.
\end{equation}
Eqs.~\eqref{eq:delta_N=4} and~\eqref{eq:cusp_N=4} are valid in the DR scheme (\ref{cp}) which preserves supersymmetry.
As ${\cal N}=4$ SYM is conformally invariant, the eigenfunctions are fixed to all orders by eq.~\eqref{loeigenf},
\beq
\Phi_{\nu n}^{{\cal N}=4}(q) = \varphi_{\nu n}(q)\,.
\eeq
Hence, $\delta_{\nu n}^{{\cal N}=4}$ is the correct NLO BFKL eigenvalue in ${\cal N}=4$ SYM.

While the NLO eigenvalue in eq.~\eqref{nloeigenv} was derived under the assumption that the eigenfunctions are the same at LO and NLO, the LO eigenfunctions \eqref{loeigenf} may themselves receive higher-order corrections in a non-conformally-invariant theory, as described by~\eqn{eq:expansion}. In fact, as the eigenvalue of a hermitian operator, the true NLO eigenvalue must be real and independent of $q^2$. $\delta_{\nu n}$ fails to meet either criterion: the right-hand side of eq.~\eqref{homognlo} depends on $q^2$ through the strong coupling constant and eq.~(\ref{nloeigenv}) contains the term $i\beta_0\,\partial_\nu \chi_{\nu n}$, which is imaginary. Note that both of these issues are absent in a conformally-invariant theory, where the strong coupling does not depend on the scale and the beta function vanishes. In particular, the term proportional to the $\beta$ function is absent in ${\cal N}=4$ SYM, as is visible in eq.~\eqref{eq:delta_N=4}, and in that case the LO eigenfunctions are indeed eigenfunctions of the NLO kernel.

In a non-conformally-invariant theory like QCD, the correct NLO eigenfunctions  are obtained through the Chirilli-Kovchegov procedure~\cite{Chirilli:2013kca,Chirilli:2014dcb}, which requires that
one constructs functions $\omega^{(1)}_{\nu n}$ and $\varphi_{\nu n}^{(1)}(q)$ such that 
\beq\label{eq:KPhi}
\left[{\cal K}^{NLO}\star \Phi_{\nu n} \right](q) = \omega_{\nu n} \Phi_{\nu n}(q)+\mathcal{O}(\overline{\alpha}_\mu^3)\,,
\eeq
with
\beq
\Phi_{\nu n} = \varphi^{(0)}_{\nu n}+\overline{\alpha}_\mu\,\varphi_{\nu n}^{(1)} \,,
\label{eq:nloeigenfn}
\eeq
and
\begin{equation}
\omega_{\nu n} =  4\,\overline{\alpha}_\mu \left( \chi_{\nu n} + \overline{\alpha}_\mu \omega_{\nu n}^{(1)} \right)\,.
\label{nloeigenvmod3}
\end{equation}

For the NLO eigenfunctions, one finds~\cite{Chirilli:2014dcb}
\begin{equation}
\Phi_{\nu n}(q) =
\varphi^{(0)}_{\nu n}(q)\left[1 + \overline{\alpha}_\mu \frac{\beta_0}2\, \ln\frac{q^2}{\mu^2}\, 
\left(  \partial_\nu {\cal P}\frac{\chi_{\nu n}}{\partial_\nu \chi_{\nu n}}
+\, i\, \ln\frac{q^2}{\mu^2}\, {\cal P}\frac{\chi_{\nu n}}{\partial_\nu \chi_{\nu n}} \right) + \mathcal{O}(\overline{\alpha}_\mu^2) \right]\,,
\label{nloeigenfexpl4}
\end{equation}
where ${\cal P}$ is the principal value prescription for $\nu=0$.

Since in a conformally-invariant theory the quantum corrections to the eigenfunctions must vanish, the quantum corrections (\ref{nloeigenfexpl4}) to the eigenfunctions are in fact proportional to the beta function. Furthermore, with the choice (\ref{nloeigenfexpl4}) of eigenfunctions, 
the NLO eigenvalue becomes~\cite{DelDuca:2017peo}
\begin{equation}
  \omega_{\nu n}^{(1)} =  \delta_{\nu n}
  - \frac{i}{2}\, \beta_0 \, \partial_\nu \chi_{\nu n} \,,
\label{nloeigenvmodcorr}
\end{equation}
which is real and independent of $q^2$, as expected. Thus, in QCD the correct NLO eigenvalue is
\beqa
\hspace{-2.5cm}  \omega_{\nu n}^{(1)} &=& 6\zeta_3 + \frac{\gamma_K^{(2)} }8\, \chi_{\nu n} 
+ \partial_\nu^2 \chi_{\nu n} -2\Phi(n,\gamma) - 2\Phi(n,1-\gamma) - \frac{1}{2}\beta_0\, \chi^2_{\nu n} \nonumber \\
  \hspace{-2.5cm} &-& 
  \frac{\Gamma(\gamma)\Gamma(1-\gamma)}{2i\nu}\left[ \psi \left( \gamma \right) - \psi \left( 1-\gamma \right) \right] \label{nloeigenvcorr} \\  
  \hspace{-2.5cm}
  &\times& \bigg[ \delta_{n0}\left( 3+ \left( 1 + \frac{N_f}{N^3_c} \right)\frac{2 + 3\gamma (1-\gamma)}{(3-2\gamma)(1+2\gamma)} \right) 
  - \delta_{|n|2}\left(\left( 1 + \frac{N_f}{N^3_c} \right)\frac{\gamma(1-\gamma)}{2(3-2\gamma)(1+2\gamma)} \right) \bigg] \,.\nonumber
\eeqa

The solution of the BFKL equation is then given by \eqn{sola} with eigenfunctions (\ref{nloeigenfexpl4}) and eigenvalue (\ref{nloeigenvmod3})
with (\ref{nloeigenvcorr}),
\beq
f(q_{1\perp},q_{2\perp},\Delta y) = \sum_{n=-\infty}^{\infty} \int_{-\infty}^{\infty} d\nu\, \Phi_{\nu n}(q_1)\,
 \Phi^\ast_{\nu n}(q_2)\, e^{\Delta y\,\omega_{\nu n}} \,.\label{eq:sola1}
\eeq
The explicit substitution of \eqn{nloeigenfexpl4} into \eqn{eq:sola1} shows that the term proportional to the $\beta$ function in \eqn{nloeigenfexpl4} can be interpreted as resetting the scale used in the strong coupling constant, such that we can use the LO eigenfunctions instead of the NLO ones provided that
we choose the scale of the strong coupling to be the geometric mean of the transverse momenta, 
$\mu^2=s_0 = \sqrt{|q_{1\perp}|^2 |q_{2\perp}|^2}$~\cite{DelDuca:2017peo},
\beq
f(q_{1\perp},q_{2\perp},\Delta y) = \sum_{n=-\infty}^{\infty} \int_{-\infty}^{\infty} d\nu\, \varphi^{(0)}_{\nu n}(q_1)\,
 \varphi^{(0) \ast}_{\nu n}(q_2)\, e^{4\,\overline{\alpha}_{\sss S}(s_0) \left( \chi_{\nu n} + \overline{\alpha}_{\sss S}(s_0) \omega_{\nu n}^{(1)}\right) \Delta y} \,.\label{eq:solafin}
\eeq

\subsection{Fourier-Mellin representation of the BFKL ladder at NLL accuracy}
\label{sec:fmnll}

We can expand the NLL part of the solution (\ref{eq:solafin}) into a power series in $\eta$ as we have done in \eqn{eqn:greensvhpl},
\beq
f^{(1)}(q_{1\bot},q_{2\bot},\eta_{s_0}) 
= \frac{1}{2\pi\, \sqrt{|q_{1\perp}|^2 |q_{2\perp}|^2} }
\,\sum_{k=1}^\infty \eta^k_{s_0}\, f^{(1)}_{k+1}(z)\,,
\label{eqn:nllsvhpl}
\eeq
with $\eta_{s_0}= 4\overline{\alpha}_{\sss S}(s_0) \Delta y$, i.e.~$\eta_{s_0}$ is given by \eqn{eq:etadef} with the scale of the strong coupling fixed at $s_0$. In \eqn{eqn:nllsvhpl}, the coefficients $f^{(1)}_{k+1}(z)$ are given by a FM transform,
\beq
f^{(1)}_{k}(z)= {\cal F}\left[\omega_{\nu n}^{(1)}\,\chi_{\nu n}^{k-2}\right] \,,
\label{eq:nllfmtransf}
\eeq
which is defined as in eq.~(\ref{eq:fmc}), with $z=-w$. Through \eqns{nloeigenv}{nloeigenvmodcorr}, we write the NLL eigenvalue (\ref{nloeigenvcorr}) in \eqn{eq:nllfmtransf} as
\beq
\omega_{\nu n}^{(1)} = 6\zeta_3 + \frac{\gamma_K^{(2)} }8\, \chi_{\nu n} - \frac{1}{2}\beta_0\, \chi^2_{\nu n}
+ \delta_{\nu n}^{(a)} + \delta_{\nu n}^{(b)} + \delta_{\nu n}^{(c)} \,, \label{nloeigenv2}
\eeq
where the first three terms, which are proportional to powers of the LO eigenvalue $\chi_{\nu n}$, are given by the FM transform (\ref{eq:fmc})
and are expressed in terms of SVHPLs as in \eqns{fkww}{fcoeffcL}. Then the coefficients in \eqn{eqn:nllsvhpl} become
\beq
\label{eq:NLL_breakdown}
	f^{(1)}_{k}(z) = 6\,\zeta_3\,f^{(0)}_{k-2}(z) + \frac{\gamma_K^{(2)}}8\,f^{(0)}_{k-1}(z) - \frac12\, \beta_0\, f^{(0)}_{k}(z)
	+ C^{(a)}_{k}(z)+ C^{(b)}_{k}(z)+ C^{(c)}_{k}(z)\,,
\eeq
where $f^{(0)}_{k}(z)$ is given in \eqns{eq:fmc}{fkww}, and we set $f_0^{(0)}(z) = {\cal F}[1] = \pi\,\delta^{(2)}(1-z)$, and with
\beq
C^{(\alpha)}_{k}(z) = {\cal F}\left[\delta_{\nu n}^{(\alpha)}\,\chi_{\nu n}^{k-2}\right]\,,
\eeq
with $\alpha=a, b, c$ and $k\ge 2$. Using \eqn{eq:delta_N=4}, in ${\cal N}=4$ SYM \eqn{eq:NLL_breakdown} becomes
\beq
\label{eq:NLLN4_breakdown}
	f^{(1){\cal N}=4}_{k}(z) = 6\,\zeta_3\,f^{(0)}_{k-2}(z) + \frac{\gamma_K^{(2)\,{\cal N}=4}}8\,f^{(0)}_{k-1}(z) 
	+ C^{(a)}_{k}(z)+ C^{(b)}_{k}(z)\,,
\eeq
with the two-loop cusp anomalous dimension in \eqn{eq:cusp_N=4}.

$C_k^{(a)}$ has the same functional form as the LL coefficients (\ref{fkww})~\cite{DelDuca:2017peo},
\beq
C_k^{(a)}(z) = \frac{|z|}{|1-z|^2}\, {\cal C}_k^{(a)}(z)\,,
\eeq
thus, ${\cal C}_k^{(a)}$ can be expressed as a linear combination of SVHPLs of uniform weight $k$ with singularities at most at $z=0$ and $z=1$.

In order to discuss $C_k^{(b,c)}(z)$, we begin by introducing
multiple polylogarithms (MPLs)~\cite{Goncharov:1998,Goncharov:2001},
which are defined as the iterated integrals,
\beq\label{eq:MPL_def}
G(a_1,\ldots,a_n;z) = \int_0^z\frac{dt}{t-a_1}\,G(a_2,\ldots,a_n;t)\,,
\eeq
except if $(a_1,\ldots,a_n) = (0,\ldots,0)$, in which case we define
\beq
G(\underbrace{0,\ldots,0}_{n\textrm{ times}};z) = \frac{1}{n!}\ln^nz\,.
\eeq
The case of harmonic polylogarithms
(HPLs)~\cite{Remiddi:1999ew} is recovered for $a_i\in \{-1,0,1\}$.
(HPLs for indices in $\{0,1\}$ are discussed in sec.~\ref{sec:SixgluonMRK}.)
In general, MPLs define multi-valued functions.
However, it is possible to consider linear combinations of MPLs such that all discontinuities cancel and the resulting function is single-valued.
A weight-1 example is the linear combination,
\beq\label{eq:SV_example}
\cG(a;z) \equiv G(a;z) + G(\overline{a};\overline{z}) = \ln\left(1-\frac{z}{a}\right) + \ln\left(1-\frac{\overline{z}}{\overline{a}}\right) = \ln\left|1-\frac{z}{a}\right|^2\,.
\eeq
The argument of the logarithm in eq.~\eqref{eq:SV_example} is positive-definite, and thus the function is single-valued. It is possible to generalize this construction to MPLs of higher weight.
In particular, in the case where the positions of the singularities $a_i$ are independent of the variable $z$, which covers the case of HPLs, one can show that there is a map ${\bf s}$ which assigns to an MPL $G(\vec a;z)$ its single-valued version ${\cal G}(\vec a;z) \equiv {\bf s}(G(\vec a;z))$. Single-valued multiple polylogarithms (SVMPLs) inherit many of the properties of ordinary MPLs. In particular, SVMPLs form a shuffle algebra and satisfy the same holomorphic differential equations and boundary conditions as their multi-valued analogues.
(See sec.~\ref{sec:SixgluonMRK} for more details for
the special case of SVHPLs.)
There are several ways to explicitly construct the map ${\bf s}$, based on the Knizhnik-Zamolodchikov equation~\cite{BrownSVHPLs,BrownSVMPLs}, the coproduct and the action of the motivic Galois group on MPLs~\cite{Brown:2013gia,Brown_Notes,DelDuca:2016lad} and the existence of single-valued primitives of MPLs~\cite{Schnetz:2016fhy}.

In \eqn{eq:NLL_breakdown}, 
$C_k^{(c)}$ can be expressed~\cite{DelDuca:2017peo}
in terms of MPLs of the type $G(a_1,\ldots,a_n;|z|)$, with $a_k\in\{-i,0,i\}$ and with
weight $0\le w\le k$. These MPLs are single-valued functions of the complex variable $z$, because the functions have no branch cut on the positive real axis. They can be re-expressed in terms of HPLs of the form $G(b_1,\ldots,b_n;|z|^2)$, with $b_i \in \lbrace -1,0 \rbrace$, and generalized inverse tangent integrals,
\begin{equation} \label{eqn:invTan}
\text{Ti}_{m_1,\dots,m_k}(|z|) = \textrm{Im}\,\text{Li}_{m_1,\dots,m_k}(\sigma_1,\ldots,\sigma_{k-1}, i\,\sigma_k\, |z|)\,,\qquad \sigma_j = \textrm{sign}(m_j)\,,
\end{equation}
where $\text{Li}_{m_1,\dots,m_k}$ denotes the sum representation of MPLs,
\beq\bsp
\text{Li}_{m_1,\dots,m_k}(z_1,\dots,z_k) &\,= \sum_{0 < n_1<n_2 < \dots < n_k} \frac{z_1^{n_1} \dots z_k^{n_k}}{n_1^{m_1}\dots n_k^{m_k}} \\
&\,= (-1)^k G\Big(\underbrace{0,\dots,0}_{m_k-1},\frac{1}{z_k},\dots,\underbrace{0,\dots,0}_{m_1-1},\frac{1}{z_1\dots z_k};1\Big)\,.
\esp\eeq

Finally we turn to $C_k^{(b)}$.  We display the two-loop result,
which is the start of a recursion in loop order $k$
based on convolution integration,
\begin{equation}\label{eq:C22}
  C_2^{(b)}(z) = \mathcal{F}\left[\delta^{(b)}_{\nu n}\right]
  = C_2^{(b,1)}(z)+C_2^{(b,2)}(z)\,,
\end{equation}
with
\beq\bsp
\label{eq:C221n2}
C_2^{(b,1)}(z) =&  \frac{|z|\,(z-\zb)}{|1+z|^2|1-z|^2}\left[\cG_{1,0}(z)-\cG_{0,1}(z)\right]\,,\\
C_2^{(b,2)}(z) =& \frac{|z|\,(1-|z|^2)}{|1+z|^2|1-z|^2}\left[\cG_{1,0}(z)+\cG_{0,1}(z)-G_{-1,0}\left(|z|^2\right)-\zeta_2\right] \,.
\esp\eeq
First, we note that $C_2^{(b)}$ is the sum of two pure functions $C_2^{(b,1)}$ and $C_2^{(b,2)}$ appearing with different rational prefactors.
Secondly, while $C_2^{(b,1)}$ is a linear combination of SVHPLs with singularities at most at $z=0$ and $z=1$, $C_2^{(b,2)}$ has a different analytic structure, with singularities also at $z=-1$. It is expressed in terms of both SVHPLs and ordinary HPLs evaluated at $|z|^2$, and it is still single-valued as a function of the complex variable $z$, because the argument of  $G_{-1,0}\left(|z|^2\right)$ is positive-definite and the function has no branch cut on the positive real axis. 

However, the single-valued polylogarithms of \eqn{eq:C221n2} do not all fall into the class of SVMPLs~\cite{BrownSVMPLs,BrownSVHPLs}, because the holomorphic derivative involves non-holomorphic rational functions. For example,
\beq\label{eq:der}
\partial_zG_{-1}\left(|z|^2\right) = \frac{1}{z+1/\zb}\,. 
\eeq
One needs then to enlarge the space of SVMPLs to a more general class of SVMPLs in one complex variable introduced by Schnetz~\cite{Schnetz:2016fhy}, with singularities at 
\beq\label{eq:loc_sing}
z=\frac{\alpha\,\zb+\beta}{\gamma\,\zb+\delta}\,,\qquad \alpha,\beta,\gamma,\delta\in\mathbb{C}\,,
\eeq
which reduce to the SVMPLs of refs.~\cite{BrownSVMPLs,BrownSVHPLs} in the case where the singularities are at constant locations. 
Since eq.~\eqref{eq:der} has a singularity at $z=-1/\zb$, we expect that the coefficients of \eqn{eq:C221n2} can be expressed in terms of Schnetz's generalized SVMPLs (gSVMPLs), $\xG(a_1,\ldots,a_n;z)$. Just like SVMPLs, gSVMPLs are single-valued, obey a shuffle algebra and vanish for
$z=0$, except if all $a_i$ are 0, in which case one has 
\beq\label{eq:init_cond}
\xG(\underbrace{0,\ldots,0}_{n\textrm{ times}};z) = \cG(\underbrace{0,\ldots,0}_{n\textrm{ times}};z) = \frac{1}{n!}\ln^n|z|^2\,.
\eeq
In addition, they satisfy the holomorphic differential equation,
\beq\label{eq:diff_eq_xG}
\partial_z\xG(a_1,\ldots,a_n;z)= \frac{1}{z-a_1}\,\xG(a_2,\ldots,a_n;z)\,,
\eeq
whose singularities are antiholomorphic functions of $z$ of the form,
\beq\label{eq:sing_def}
a_ i = \frac{\alpha\,\zb+\beta}{\gamma\,\zb+\delta}\,,\textrm{ for some } \alpha,\beta,\gamma,\delta\in\mathbb{C}\,.
\eeq

One can write $C_{k}^{(b)}(z)$ in the form~\cite{DelDuca:2017peo},
\beq\bsp\label{eq:C2k_to_cC}
C_{k}^{(b)}(z) =& \; \frac{|z|\,(z-\zb)}{|1+z|^2|1-z|^2}\cC^{(b,1)}_k + \frac{|z|\,(1-|z|^2)}{|1+z|^2|1-z|^2}\cC^{(b,2)}_k\,,
\esp\eeq
where the functions $\cC^{(b,i)}_k$ have uniform weight $k$, which at two and three loops can be expressed in terms of SVHPLs and ordinary HPLs evaluated at $|z|^2$, while starting from four loops they are expressed in terms of gSVMPLs with $a_i\in \{-1,0,1,-1/\zb\}$.

\subsection{Transcendental weight of the BFKL ladder at NLL accuracy}
\label{sec:weightnll}

Collecting the results of sec.~\ref{sec:fmnll}, we see that the
perturbative expansion coefficients $f_{k}^{(1)}$
of the BFKL solution at NLL accuracy (\ref{eqn:nllsvhpl}) for
both QCD and ${\cal N}=4$ SYM can be expressed in terms of single-valued polylogarithms, which range from SVHPLs to gSVMPLs.
In ${\cal N}=4$ SYM, the single-valued polylogarithms of \eqn{eq:NLLN4_breakdown} have a uniform and maximal transcendental weight $k$.
In QCD, the single-valued polylogarithms of \eqn{eq:NLL_breakdown}
have weight up to $k$.
The weight drop in QCD occurs because the beta function term has weight $k-1$,
the cusp anomalous dimension terms in \eqn{eq:k2} have weight zero and two,
and so the corresponding terms in \eqn{eq:NLL_breakdown}
have weight $k-2$ and $k$, and finally the terms due to $C_{k}^{(c)}$ have weight $0\le w\le k$.  Thus the QCD result is not a maximal weight function.
Furthermore, since $C_{k}^{(c)}$ has terms of weight $k$, i.e.~of maximal weight,
and it is missing in \eqn{eq:NLLN4_breakdown}, the maximal weight 
of \eqn{eq:NLL_breakdown} cannot coincide with \eqn{eq:NLLN4_breakdown}.
That is, the maximal weight terms of the QCD color-singlet BFKL ladder
in momentum space at NLL accuracy do {\it not} match one by one the terms of the
color-singlet ladder in ${\cal N}=4$ SYM. In contrast, the anomalous dimensions
of the leading-twist operators which control Bjorken scaling violation
have a uniform and maximal transcendental weight in ${\cal N}=4$ SYM,
which also matches the maximal weight part of the corresponding anomalous
dimensions in QCD, once one sets
$C_F\to C_A$~\cite{Kotikov:2000pm,Kotikov:2002ab,Kotikov:2003fb}.

Consider the color-singlet BFKL ladder in a generic $SU(N_c)$ gauge theory
with scalar or fermionic matter in arbitrary representations.
Is there any other theory where the momentum-space results have a uniform and
maximal weight which also agrees with the maximal weight part of the
BFKL ladder in QCD at NLL accuracy~\cite{DelDuca:2017peo}?
In a theory where the gauge group is minimally coupled to matter,
the BFKL eigenvalue at NLL accuracy is determined entirely by the gauge group and matter content of the theory~\cite{Kotikov:2000pm}, but it is independent of the details of the other interactions in the theory, like the Yukawa couplings between the fermions and the scalars, which would start occurring only at higher accuracy. As a consequence, we can repeat the analysis of the transcendental weight properties for generic gauge theories as a function of the fermionic and scalar matter content of the theory.
It was then found~\cite{DelDuca:2017peo} that the necessary and sufficient conditions for a theory to have a BFKL ladder at NLL accuracy of uniform transcendental weight in momentum space are that:
\begin{enumerate}
\item the one-loop beta function vanishes;
\item the two-loop cusp anomalous dimension is proportional to $\zeta_2$;
\item the contribution from the $n=\pm 2$ term of $\delta^{(c)}_{\nu n}$ in eq.~(\ref{nloeigenv3}) vanishes.
\end{enumerate}
Since the last condition contains terms of maximal weight, we conclude that there is no theory such that the BFKL ladder at NLL accuracy has uniform and maximal weight and agrees with the maximal weight terms in QCD.

In particular, for theories with matter only in the fundamental and adjoint representations, the necessary conditions for a gauge theory to have a BFKL ladder at NLL accuracy of uniform and maximal transcendental weight were established~\cite{DelDuca:2017peo}.  For theories with the maximal weight property, the field content can be arranged into supersymmetric multiplets, although supersymmetry was not an input to the analysis. 
In particular, in addition to ${\cal N}=4$ SYM, only three theories were found to satisfy the constraints. They are ${\cal N}=2$ superconformal QCD with $N_f=2N_c$ hypermultiplets~\cite{Gadde:2009dj}; ${\cal N}=1$ super-QCD with $N_f=3N_c$ flavors in the fundamental representation; and an ${\cal N}=1$ solution with two flavors in the adjoint and $N_c$ flavors in the fundamental representations.

\subsection{Toward a BFKL ladder at NNLL accuracy}
\label{sec:nnll}

\begin{figure}
  \centerline{\includegraphics[width=0.7\columnwidth]{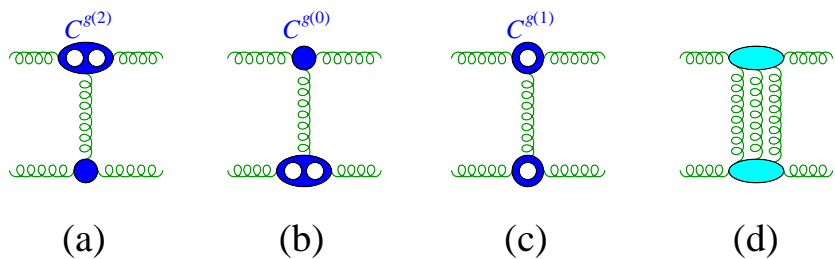} }
  \caption{Additional contributions to the two-loop four-gluon amplitude at NNLL accuracy.
  (a, b) The two-loop impact factor is represented by the twice pierced blue blob.
  (c) The one-loop impact factor squared.
  (d) The three-Reggeized-gluon exchange.}
\label{fig:twoloopnnll}
\end{figure}

Beyond NLL accuracy, gluon Reggeization~\cite{Fadin:2006bj,Fadin:2015zea} and Regge pole factorization break down~\cite{DelDuca:2001gu}. The real part of $2\to 2$ amplitudes is not anymore given only by the exchange of a Reggeized gluon, as in \eqn{eq:reamp4gnll}, corresponding to a Regge pole in the complex angular momentum plane. It also involves contributions from the exchange of three Reggeized gluons~\cite{DelDuca:2001gu,DelDuca:2011wkl,DelDuca:2011ae,DelDuca:2013ara,DelDuca:2014cya,Fadin:2016wso,Caron-Huot:2017fxr,Fadin:2017nka,Falcioni:2020lvv,Falcioni:2021buo,Falcioni:2021dgr}, corresponding to a Regge cut in the angular momentum plane. Accordingly, in the non-logarithmic term of the two-loop four-gluon amplitude, $M^{(-,2,0)}_{4g}$, which has NNLL accuracy,
both the two-loop impact factor, fig.~\ref{fig:twoloopnnll}(a, b), and the three-Reggeized-gluon exchange, fig.~\ref{fig:twoloopnnll}(d), contribute and mix up, invalidating Regge pole factorization at the $N_c$-subleading level~\cite{DelDuca:2001gu}.
In order to disentangle those two contributions, a prescription based on infrared factorization~\cite{DelDuca:2011wkl,DelDuca:2011ae} has been introduced~\cite{DelDuca:2013ara,DelDuca:2014cya}, which identifies the usual diagonal terms of the color octet exchange with the two-loop impact factor and the non-diagonal ones with the factorization-violating terms.
An analogous prescription, based on the explicit computation of the NNLL corrections ${\cal M}_{4g}^{(-,n,n-2)}$ to the four-gluon amplitude~\cite{Falcioni:2020lvv,Falcioni:2021buo,Falcioni:2021dgr}, restricts the planar multi-Reggeon contributions to occur only at two and three loops. These contribute to the Regge pole and may be factorized together with the Reggeized gluon as in~\eqn{sudall}, while the non-planar multi-Reggeon contributions make up the Regge cut.
Making the above prescriptions explicit to three loops,
one can predict how the factorization-violating terms propagate into the
single-logarithmic term of the three-loop amplitude $M_{4g}^{(-,3,1)}$,
and thus have an operative way to disentangle the factorization-violating terms from the three-loop gluon Regge trajectory~\cite{DelDuca:2021vjq,Falcioni:2021dgr,Caola:2021izf}.

The possibility of disentangling terms based on the exchange of one Reggeized gluon
from factorization-violating terms hints that the BFKL equation,
which is based on the exchange of one Reggeized gluon, may be extended to NNLL accuracy.
In addition, there are reasons, based on the integrability of amplitudes in MRK in the large $N_c$ limit~\cite{Lipatov:2009nt},
to believe that Regge pole factorization will be simpler in that case.
This warrants an analysis of the terms which would contribute to the BFKL equation at NNLL accuracy.

\begin{figure}
  \centerline{\includegraphics[width=0.8\columnwidth]{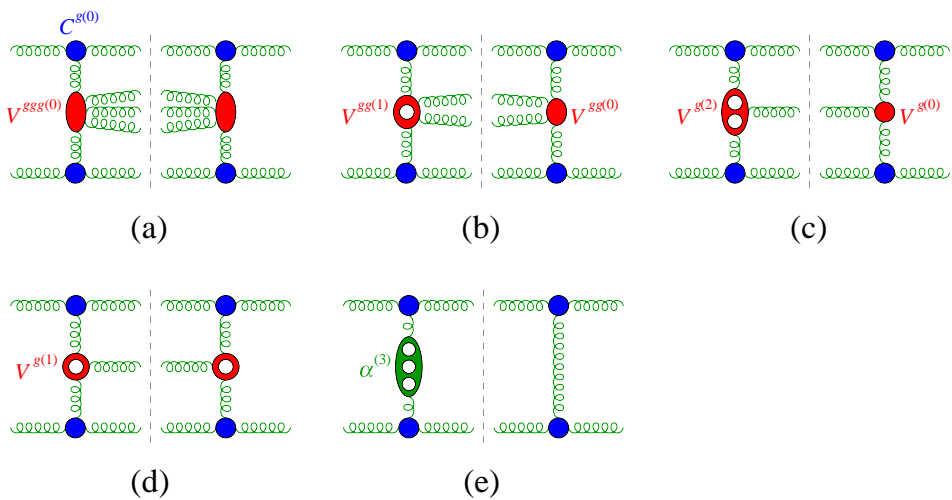} }
  \caption{(a) The red blob along the gluon ladder represents the three-gluon central-emission vertex within the tree seven-gluon amplitude.
  (b) The pierced red blob represents the one-loop two-gluon central-emission vertex within the one-loop six--gluon amplitude.
  (c) The twice pierced red blob represents the two-loop central-emission vertex within the two-loop five--gluon amplitude.
  (d) The square of the one-loop five-gluon amplitude in MRK.
  (c) The thrice pierced green blob represents the three-loop gluon Regge trajectory within the three-loop four--gluon amplitude.}
\label{fig:nnllbfkl}
\end{figure}

At NNLL accuracy, the kernel of the BFKL equation will have contributions
from the CEV for the emission of three partons along the gluon
ladder~\cite{DelDuca:1999iql,Antonov:2004hh,Duhr:2009uxa}, 
evaluated in next-to-next-to-multi-Regge kinematics (NNMRK), fig.~\ref{fig:nnllbfkl}(a),
\beq
\ldots \gg p_4^+\simeq p_5^+ \simeq p_6^+ \gg \ldots\,;
\label{eq:nnmrk}
\eeq
from the one-loop corrections to the CEV for the emission of
two gluons~\cite{Byrne:2022wzk} or of a quark-antiquark pair along the
gluon ladder, evaluated in the NMRK of \eqn{eq:nmrk45},
fig.~\ref{fig:nnllbfkl}(b);
from the two-loop corrections to the single-gluon CEV in MRK,
fig.~\ref{fig:nnllbfkl}(c);
and from the square of the one-loop five-gluon amplitude in MRK,
which will contain the square of
one-loop corrections to the single-gluon CEV, fig.~\ref{fig:nnllbfkl}(d).
Once those contributions are assembled into the kernel at NNLL accuracy,
the infrared divergences of the kernel must cancel the divergences of
the three-loop Regge trajectory, fig.~\ref{fig:nnllbfkl}(e).
Carrying out all the phase space integrations is a challenge for the future.

%%%%%%%%%%%%%%%%%%%%%%%%%%%%%%%%%%%%%%%%%%%%%%%%%%%%%%%%%%%%%%%%%%

\section{The multi-Regge limit of planar ${\cal N}=4$ SYM amplitudes}
\label{sec:planarN4regge}

\subsection{Overview}
\label{sec:planarN4overview}

The maximally supersymmetric cousin of QCD is ${\cal N}=4$ super-Yang-Mills
theory. Instead of having $n_f$ flavors of quarks in the fundamental
representation, its matter content consists of 4 fermions (gluinos)
$\Gamma_A$ and 6 real scalars $S_{AB}$
in the adjoint representation of the gauge group $G$.
The four supersymmetries can be used to package all $2^4 = 16$ helicity
states into a single {\it on-shell superfield},
\beq
\Phi(k,\eta) = G^+ +\eta^A\Gamma_A+\tfrac{1}{2!}\eta^A\eta^B S_{AB}+\tfrac{1}{3!}\eta^A\eta^B\eta^C\varepsilon_{ABCD}\bar \Gamma^D
+\tfrac{1}{4!}\eta^A\eta^B\eta^C\eta^D\varepsilon_{ABCD}G^- ,
\label{eq:superfield}
\eeq
where $\eta^A$ is a Grassmann variable with $A=1,2,3,4$,
and $\varepsilon_{ABCD}$ is the Levi-Civita antisymmetric tensor.

The large amount of supersymmetry leads to the vanishing of ultraviolet
divergences, so the beta function vanishes, $\beta(g)=0$, and
the theory is conformal~\cite{Mandelstam:1982cb,Brink:1982wv,Howe:1983sr}.
If we further take the gauge group to be
$G = SU(N_c)$ and send $N_c \to \infty$
--- the large $N_c$ or planar limit ---
then only planar Feynman diagrams
contribute~\cite{tHooft:1973alw}, and the 't Hooft coupling constant
is defined as
\beq
g^2 \, \equiv \, \frac{\lambda}{16\pi^2}
\, \equiv \, \frac{N_c \gs^2}{16\pi^2}
\, = \frac{N_c \as}{4\pi} \,.
 \label{eq:tHooft}
\eeq
We generally take $N_c \to \infty$ holding $\lambda$ or $g^2$ fixed.
In this limit, the color-decomposition of $n$-gluon amplitudes simplifies
drastically, to a sum over non-cyclic permutations of single-trace terms,
\beqa
{\cal A}_n(a_1,a_2,\ldots,a_n) &=& \gs^{n-2}
\sum_{\ell=0}^\infty g^{2\ell} \sum_{\sigma\in S_n/Z_n} 
    {\rm Tr}(T^{a_{\sigma(1)}} T^{a_{\sigma(2)}} \cdots T^{a_{\sigma(n)}})\nn\\
&&\null \times A_n^{(\ell)}(\sigma(1),\sigma(2),\ldots,\sigma(n))
+ {\cal O}\biggl(\frac{1}{N_c^2}\biggr).
\label{eq:colordecomp}
\eeqa
Multi-trace contributions in \eqn{eq:colordecomp} are suppressed by
at least one power of $1/N_c^2$ in the color-summed cross section.
Interferences between different non-cyclic orderings $\sigma$ are also
color-suppressed, so the $n$ gluons can be taken to have
a definite cyclic ordering, by default $1,2,\ldots,n$.

In the planar limit, ${\cal N}=4$ SYM
becomes integrable~\cite{Beisert:2010jr},
giving rise to additional symmetries and the prospect of solving
for scattering amplitudes exactly at finite coupling.
In the limit of large $\lambda$, AdS/CFT duality implies
that scattering amplitudes can be computed in terms of minimal
area surfaces in anti-de Sitter space,
whose boundary is a closed light-like polygon~\cite{Alday:2007hr}.
This picture led to the duality between amplitudes
and polygonal Wilson loops
at arbitrary coupling~\cite{Alday:2007hr,Brandhuber:2007yx,Drummond:2007cf,
  Drummond:2007au,Bern:2008ap,Drummond:2008aq,Alday:2008yw,Adamo:2011pv},
which also implies invariance under a set of {\it dual conformal} transformations,
which are distinct from, and in addition to,
ordinary (position space) conformal symmetry.

These five additional dual conformal symmetries make
the kinematic dependence of the four- and five-gluon amplitudes trivial.
To all loop orders they are given by the Bern-Dixon-Smirnov (BDS)
ansatz~\cite{Bern:2005iz}, which is essentially the exponential
of the one-loop amplitude multiplied by the light-like
cusp anomalous dimension $\Gc$, which is known to all orders
in $g^2$~\cite{Beisert:2006ez}.
Thus the multi-loop structure of four- and five-gluon amplitudes
in MRK is rather simple in planar ${\cal N}=4$ SYM: it is governed by the
behavior of the one-loop amplitude and the constants in the BDS ansatz.

Starting with the six-gluon amplitude, the structure
becomes much richer.  In general kinematics, there are three
independent dual conformally invariant cross ratios for six-gluon scattering.
In an appropriate MRK limit, one of these three variables generates large
logarithms in the rapidity. The coefficients of the large logarithms
depend on two remaining variables parametrizing the complexified transverse
momenta. The two variables lie on a complex sphere
$(z,\zb)=(-w,-\ws)$ with three punctures, at $0,1,\infty$,
and the coefficients are real-analytic functions of $(z,\zb)$,
which means that they are single-valued around all three punctures.
For higher-point amplitudes, the general picture is similar.
Each additional gluon adds three more variables; one is a large
logarithm and two more are complex pairs.  The set of punctures
becomes much richer because it includes limits where different variables
approach each other, as well as where they approach the three punctures.

The prospect of a finite-coupling solution for the scattering of more than
five gluons has already been realized in various kinematical limits.
For example, the six-gluon amplitude is dual to a hexagonal Wilson loop.
The limit that two gluons become collinear is conformally equivalent
to pulling apart the hexagon into two halves, separated by a long
distance.  In this limit, the operator product expansion (OPE) is
dominated by low-twist operators~\cite{Alday:2010ku}, or flux-tube
excitations, whose anomalous dimensions have been computed at finite
coupling~\cite{Basso:2010in}.
The interactions of these excitations are characterized by an
integrable two-dimensional $S$ matrix, which is closely related
to certain pentagon transitions.  A general $n$-gluon amplitude,
or Wilson $n$-gon, can be tiled by $(n-4)$ pentagons, and the resulting
finite-coupling formula for the (multi) near-collinear limit
is referred to as the pentagon OPE or flux-tube
expansion~\cite{Basso:2013vsa,Basso:2013aha,Basso:2014koa,Basso:2014nra}.

In the six-gluon case, the contribution of a single flux-tube excitation
was analytically continued to give an all-orders proposal
for the multi-Regge limit~\cite{Basso:2014pla}, which
encompasses both the (adjoint) BFKL eigenvalue and the impact factor.
More recently, a similar analysis was used to propose an all-orders
formula for the central-emission vertex, which first enters
seven-gluon MRK, and should allow an MRK description at arbitrary
multiplicity and coupling.  In the remainder of this section
we develop these ideas further.

\subsection{Six-gluon MRK}
\label{sec:SixgluonMRK}

Dual conformal symmetry means that suitably normalized amplitudes
have a full $SO(4,2)$ invariance that acts in momentum space, or more precisely
on dual coordinates $x_i$, instead of the usual
Poincar\'e invariance ($SO(3,1)$ Lorentz symmetry plus four translations).
The dimension of $SO(4,2)$ is ${6 \choose 2} = 15$, while the Poincar\'e
group has ${4 \choose 2} + 4 = 10$ generators.  The additional five
dual conformal symmetry generators consist of dilatations and four
special (dual) conformal generators.  The latter can be represented
as an inversion $x_i^\mu \to x_i^\mu/x_i^2$, followed by
an infinitesimal translation
$x_i^\mu \to x_i^\mu + \epsilon^\mu$, followed by another inversion.
Here the {\it dual coordinates} $x_i^\mu$ are the corners of the light-like
polygon, so their differences are the momenta $p_i^\mu$:
\beq
x_i^\mu - x_{i+1}^\mu = p_i^\mu \,, \qquad
x_{ij}^2 = (p_i + p_{i+1} \cdots + p_{j-1})^2 \,.
\label{eq:xprelation}
\eeq
Under an inversion,
\beq
x_{ij}^2 \to \frac{x_{ij}^2}{x_i^2 x_j^2} \,.
\label{eq:inversion}
\eeq
Dual conformally invariant functions are functions $f(u_{ijkl})$
of the cross ratios
\beq
u_{ijkl} \equiv \frac{x_{ij}^2 x_{kl}^2}{x_{ik}^2 x_{jl}^2} \,,
\label{crossratios}
\eeq
because under the inversion~(\ref{eq:inversion}) the factors
of $x_i^2 x_j^2 x_k^2 x_l^2$ cancel between numerator and denominator.

In four dimensions, the number of kinematic variables for a
Poincar\'e-invariant $n$-point process is $3n-10$: $3n$ for the spatial
momentum components of the $n$-particles (which determine the energies),
minus 10 for the symmetries. This formula gives 2 for the 4-point process,
namely the familiar Mandelstam variables $s$ and $t$ (with $u = -s-t$ in
the massless case), and 5 for the 5-point process, which could be
taken to be $s_{i,i+1} = (p_i + p_{i+1})^2$, $i=1,2,\ldots,5$.
The five additional symmetry generators of dual conformal invariance reduce
the number of kinematic variables by five more, to $3n-15$.
There are no variables left for $n=4$ or 5, and the (MHV) amplitudes
are given by the BDS ansatz~\cite{Bern:2005iz},
\beq
A_n^{\rm BDS} = A_n^{(0)} \, \text{exp}
\left[ \sum_{\ell=1}^{\infty} g^{2\ell}
  \left( f^{(\ell)}(\e) M_{n}^{(1)}(\ell\e) + C^{(\ell)} \right)
  \right] \,,
\label{eq:BDS}
\eeq
where $A_n^{(0)}$ is the tree-level Parke-Taylor amplitude~\cite{Parke:1986gb},
$M_{n}^{(1)}(\e)$ is the one-loop MHV amplitude
(divided by the tree)~\cite{Bern:1994zx},
\begin{equation}
\label{BDSf}
f^{(\ell)}(\e) = f_{0}^{(\ell)} + \e \, f_{1}^{(\ell)} + \e^2 \, f_{2}^{(\ell)}
\end{equation}
is a set of three constants,
and $C^{(\ell)}$ is an additional, finite constant.
Here $f_{0}$ is the cusp anomalous dimension,
known to all orders in perturbation theory~\cite{Beisert:2006ez},
\beq
  f_{0}(g^2) \, \equiv \, \Gamma(g^2) \, = \, \frac{\Gc(g^2)}{4}\, =\,
g^2 - 2\,\zeta_2 \, g^4 + 22 \, \zeta_4 \, g^6 
             - ( 219 \, \zeta_6 + 8 \, \zeta_3^2 ) \, g^8
             + \cdots \,.
\label{eq:gamma_cusp}
\eeq
The convention for normalizing the cusp anomalous dimension
in planar ${\cal N}=4$ differs from that in QCD by a factor of two:
\beq
\gamma_K \ =\ 2 \, \Gamma_{\rm cusp} \,.
\label{eq:twocuspnorms}
\eeq
In \eqn{BDSf}, $f_{1}^{(\ell)}$ is related to the gluon collinear
anomalous dimension defined in \eqn{eq:collad} by
\beq
f_1^{(\ell)}\ =\ - \frac{\ell}{2} \, \gamma_g^{(\ell)\, {\cal N}=4}
\ = \ \frac{\ell}{4} {\cal G}_0^{(\ell)\, {\cal N}=4} \,,
\label{eq:f1relation}
\eeq
where the second definition is used in e.g.~ref.~\cite{Dixon:2017nat}.
It is known to four loops~\cite{Cachazo:2007ad,Dixon:2017nat,Agarwal:2021zft}.
The finite constants $f_{2}^{(\ell)}$ and $C^{(\ell)}$ are known (numerically)
to three loops~\cite{Bern:2005iz,Spradlin:2008uu}.

The BDS ansatz provides the complete amplitude for $n=4$ and 5
because it is the unique solution
to an anomalous dual conformal Ward identity~\cite{Drummond:2007cf}.
Starting at $n=6$, the solution is not unique, because of the existence
of three dual conformal cross ratios,
\beq
u_1 = \frac{x_{13}^2x_{46}^2}{x_{14}^2x_{36}^2} = \frac{s_{12}s_{45}}{s_{123}s_{345}}
\,, \qquad
u_2 = \frac{x_{24}^2x_{51}^2}{x_{25}^2x_{41}^2} = \frac{s_{23}s_{56}}{s_{234}s_{123}}
\,, \qquad
u_3 = \frac{x_{35}^2x_{62}^2}{x_{36}^2x_{52}^2} = \frac{s_{34}s_{61}}{s_{345}s_{234}}
\,.
\label{eq:ui}
\eeq
The full six-point amplitude can be written as
\beq
A_6^{\rm MHV}(s_{i,j},\e) = A_6^{\rm BDS}(s_{i,j},\e)
\, \exp\Bigl[ R_6(u_1,u_2,u_3) \Bigr] \,,
\label{eq:Rdef}
\eeq
where $R_6$ is the {\it remainder function}.
The six-point one-loop amplitude entering the BDS ansatz is
\beqa
\hspace{-2.5cm}
M_6^{(1)} &=& \hat{M}_6^{(1)} \, + \, {\cal E}_6^{(1)}(u_i) \,, \label{eq:M6def}\\
\hspace{-2.5cm}
\hat{M}_6^{(1)} &=& \sum_{i=1}^6 \biggl[
  - \frac{1}{\e^2} + \frac{\ln(-s_{i,i+1})}{\e}
  - \ln(-s_{i,i+1}) \Bigl( \ln(-s_{i+1,i+2})
  - \frac{1}{2} \ln(-s_{i+3,i+4}) \Bigr) \biggr]
+ 6 \zeta_2 \,, \nn
\eeqa
where
\beq
{\cal E}_6^{(1)} = \sum_{i=1}^3 {\rm Li}_2\biggl(1-\frac{1}{u_i}\biggr) \,.
\label{eq:E6def}
\eeq
It is also possible to normalize by a BDS-like
ansatz~\cite{Alday:2009dv,Dixon:2015iva}, which uses $\hat{M}_6^{(1)}$
instead of $M_6^{(1)}$ in \eqn{eq:BDS},
thus omitting the finite, dual conformally invariant part
${\cal E}_6^{(1)}$ of the one-loop amplitude.
This normalization leads to improved causal properties for
bootstrapping at generic kinematics; namely, the Steinmann relations
remain manifest~\cite{Caron-Huot:2016owq}.
However, for discussing MRK the remainder function $R_6$ is more suitable,
because it vanishes in all collinear and soft limits, and a soft limit is
equivalent to a Euclidean version of MRK.

There are various possible six-point MRK limits, but the one that gives
rise to the most interesting behavior~\cite{Bartels:2008ce}
is the case of $2\to4$ scattering when the two
incoming gluons are as far as they can get 
from each other in the cyclic color ordering.
For the cyclic color ordering $1,2,\ldots,6$, we take
gluons 3 and 6 to be incoming and gluons $1,2,4,5$ outgoing.
(This configuration is sometimes also described by starting with
the process $1+2 \to 3+4+5+6$, and moving legs 4 and 5 into the initial
state, between legs 1 and 2.)
The strong rapidity ordering for MRK, in terms of color-ordered
Mandelstam variables, is
\beq
s_{12} \gg s_{345},s_{123} \gg s_{34},s_{45},s_{56} \gg s_{23},s_{61},s_{234} \,.
\label{eq:MRK36}
\eeq
In terms of the cross ratios~(\ref{eq:ui}), the limit is
\beq
u_1 \to 1, \qquad u_2 \to 0, \qquad u_3 \to 0,
\label{eq:uiMRKlim}
\eeq
with $u_2/(1-u_1)$ and $u_3/(1-u_1)$ held fixed.
It is important to note that while $s_{12}$ and $s_{45}$ are
positive, the rest of the invariants listed in \eqn{eq:MRK36}
are negative.  These sign assignments in \eqn{eq:ui}
lead to $u_1,u_2,u_3>0$.

There is an unphysical Euclidean branch, or Riemann sheet,
where all Mandelstam invariants are spacelike, and so $u_1,u_2,u_3>0$
there as well.
On the Euclidean sheet, all loop integrals are real, and hence the
scattering amplitude is real.  The $2\to4$ sheet for physical
Minkowski scattering is on a different Riemann sheet from the
Euclidean sheet, despite also having $u_1,u_2,u_3>0$.
(Physical $2\to4$ scattering also requires $u_1,u_2,u_3<1$ and
$\Delta<0$, where $\Delta$ is given in
\eqn{eq:Deltadef}~\cite{Byrne:2022wzk}.)
Because $u_1$ in \eqn{eq:ui} contains two positive (timelike)
invariants in the numerator, and two negative (spacelike, or Euclidean)
invariants in the denominator, it has to be continued by wrapping once
around the complex origin,
\beq
u_1 \to u_1 e^{-2\pi i} \,.
\label{eq:u1continue}
\eeq
On the other hand, $u_2$ and $u_3$ are composed entirely of spacelike
invariants, so they do not have to be continued at all from the
Euclidean sheet.
The analytic continuation~(\ref{eq:u1continue}) generates discontinuities
that diverge in MRK, even though the remainder function $R_6$
vanishes in the same limit~(\ref{eq:uiMRKlim}) on the Euclidean sheet.

The finite,
dual conformally invariant part of six-gluon scattering amplitudes
in planar ${\cal N}=4$ SYM in the ``bulk'', i.e.~for arbitrary kinematics,
are built from a polylogarithmic function space described by
nine letters:
\beq
{\cal S}_{\rm hex} = \{ u_i \,, 1-u_i \,, y_i \}, \quad i = 1,2,3,
\label{eq:Shex}
\eeq
where
\beqa
y_i &=& \frac{u_i-z_+}{u_i-z_-} \,, \qquad
z_\pm = \frac{ -1 + u_1 + u_2 + u_3 \pm \sqrt{\Delta}}{2} \,,
\label{eq:yidef} \\
\Delta &=& (1-u_1-u_2-u_3)^2-4u_1u_2u_3 \,.
\label{eq:Deltadef}
\eeqa
As reviewed in Chapter 5 of this SAGEX Review~\cite{Papathanasiou:2022lan},
the meaning of the symbol alphabet ${\cal S}$
\cite{Goncharov:2010jf} is that every function $F$ in
the corresponding function space has a ``{\it d}\,log'' derivative
structure with a finite number of terms,
\beq
dF = \sum_{s_k \in {\cal S}} F^{s_k} \ d\ln s_k \,,
\label{eq:dFcop}
\eeq
where $s_k$ are the symbol letters, 
and if $F$ has weight $n$ then $F^{s_k}$ has weight $n-1$.\footnote{
See ref.~\cite{Duhr:2014woa} for an introduction to polylogarithms,
the symbol, and all that.}

The $y_i$ letters are fairly complicated in the bulk, at least
in terms of the cross-ratios $u_i$.  (All 9 letters can be parametrized
rationally in the bulk using the $y_i$ variables~\cite{Dixon:2011pw},
or momentum twistors, which are associated with the
Grassmannian ${\rm Gr}(4,6)$, see e.g.~ref.~\cite{Caron-Huot:2020bkp},
or the pentagon OPE parametrization in
refs.~\cite{Basso:2013vsa,Basso:2013aha}.)

Here we only need to parametrize the MRK limit~(\ref{eq:uiMRKlim}).
It is convenient to take
\beq
\frac{u_2}{1-u_1} = \frac{1}{(1-z)(1-\zb)} \,, \qquad
\frac{u_3}{1-u_1} = \frac{z\zb}{(1-z)(1-\zb)} \,,
\label{eq:zdef}
\eeq
as this rationalizes $\sqrt\Delta \approx (1-u_1)(z-\zb)/|1-z|^2$.
Thus the $y_i$ become rational too,
\beq
y_1 \to 1 \,, \qquad y_2 \to \frac{1-\zb}{1-z} \,, \qquad
y_3 \to \frac{(1-z)\zb}{(1-\zb)z} \,,
\label{eq:yiz}
\eeq
and they are pure phases when $\zb$ is the complex conjugate of $z$.

It is apparent from \eqns{eq:zdef}{eq:yiz} that the
bulk symbol alphabet~(\ref{eq:Shex}) collapses in MRK
to the four letters
\beq
{\cal S}_{\rm hex,\, MRK} = \{ z, 1-z, \zb, 1-\zb \} \,.
\label{eq:ShexMRK}
\eeq
There is also the infinitesimal letter $(1-u_1)$, but it just corresponds
to free powers of $\ln(1-u_1)$.
Using the characterization~(\ref{eq:dFcop}), the derivatives
of any function in this space have the form
\beq
\frac{\partial F}{\partial z} = \frac{F^z}{z} - \frac{F^{1-z}}{1-z} \,,
\qquad
\frac{\partial F}{\partial \zb} = \frac{F^{\zb}}{\zb}
                                - \frac{F^{1-\zb}}{1-\zb} \,.
\label{eq:dFzzb}
\eeq
These relations imply that the relevant function space
must be built from HPLs~\cite{Remiddi:1999ew} in $z$ and $\zb$.

HPLs with indices $\{0,1\}$ are defined for
binary strings $\vec{w}$ with elements $w_k \in \{0,1\}$.
They are defined iteratively by
\beq
H_{0,\vec{w}}(z) = \int_0^z \frac{dt}{t} H_{\vec{w}}(t),
\qquad
H_{1,\vec{w}}(z) = \int_0^z \frac{dt}{1-t} H_{\vec{w}}(t),
\label{eq:Hzdef}
\eeq
as well as the initial condition $H(z)=1$ for the null string,
and the special case of all zeroes,
\beq
H_{\vec{0}_n}(z) = \frac{1}{n!} \ln^n z \,.
\label{eq:H0zdef}
\eeq
The {\it weight} of $H_{\vec{w}}$ is the number of integrations, or the length
$|\vec{w}|$ of the binary string $\vec{w}$.
HPLs obey the differential relations,
\beq
\frac{\partial H_{0,\vec{w}}(z)}{\partial z} = \frac{H_{\vec{w}}(z)}{z} \,,
\qquad
\frac{\partial H_{1,\vec{w}}(z)}{\partial z} = \frac{H_{\vec{w}}(z)}{1-z} \,,
\label{eq:Hdiffdef}
\eeq
which matches the structure of \eqn{eq:dFzzb}.
Thus the relevant MRK function space involves a tensor product of
the singular logarithm, HPLs in $z$, and HPLs in $\zb$:
\beq
{\cal F}_{\rm MRK} \subset 
\left\{ \ln^k(1-u_1),\ k\ge0 \right\} 
\otimes \left\{ H_{\vec{w}}(z),\ w_k\in\{0,1\} \right\}
\otimes \left\{ H_{\vec{w}}(\zb),\ w_k\in\{0,1\} \right\} \,.
\label{eq:FnsMRK}
\eeq
At weight $n$, there are $2^n$ possible HPLs with indices $\{0,1\}$.
They obey a {\it shuffle algebra}, so that
\begin{equation}
\label{eq:HPL_shuffle}
H_{\vec{w}_1}(z)\,H_{\vec{w}_2}(z)
= \sum_{w\in w_1 \ssha w_2} H_{\vec{w}}(z)\,,
\end{equation}
where ${w_1}\sha{w_2}$ is the set of shuffles,
or mergers of the sequences $w_1$ and $w_2$
that preserve their individual orderings. Equation~(\ref{eq:HPL_shuffle})
can be used, from right to left, to express the $2^n$ functions
in terms of sums
and products of a much smaller set of functions, which is
called the {\it Lyndon} basis, because there is one $H_{\vec{w}_L}$ for each
Lyndon word.  Lyndon words $w_L$, when split into any two sub-words
$u$ and $v$, always have $u<v$ lexicographically.
The number of binary Lyndon words for weight
$1,2,3,\ldots$ is $2,1,2,3,6,9,18,30,\ldots$. The first few Lyndon words
are $w_L = 0, 1; 01; 001, 011; 0001, 0011, 0111$.
The {\it linear} basis is obtained by applying \eqn{eq:HPL_shuffle}
repeatedly from left to right, so that there are no more products.

\Eqn{eq:FnsMRK} is not the whole story;
there is one more restriction, single-valuedness,
which is related to the location of branch cuts.
On the Euclidean sheet, for massless scattering amplitudes,
branch cuts originate when Mandelstam variables vanish, since that is
where physical scattering can first turn on. Given \eqn{eq:ui},
these cuts can only originate at $u_i=0$ or $u_i=\infty$.
In the symbol, branch cut locations correspond to the vanishing of
first entries (or their inverses), leading to the
{\it first-entry condition}~\cite{Gaiotto:2011dt} which for
general six-point kinematics states that only the symbol letters $u_i$
in \eqn{eq:Shex} are allowed to occupy the first entry in any term
in the symbol, not $1-u_i$ nor $y_i$.  In fact, the $y_i$ letters
do not actually appear until the third entry, because of integrability
(equality of mixed partial derivatives).
When the analytic continuation~(\ref{eq:u1continue}) is performed,
at symbol level it corresponds to clipping off a $u_1$ from the front of
the symbol, exposing the second entry.  Because this entry is
always a $u_i$ or $1-u_i$, as one takes the limit~(\ref{eq:zdef}),
the first entry always involves the {\it pairs} $z\zb$ and
$(1-z)(1-\zb)$.\footnote{%
There is a subtlety related to contributions
from higher discontinuities, which requires demonstrating
that clipping an arbitrary number of $u_1$'s from the front
of the symbol never exposes a $y_i$~\cite{Dixon:2021nzr}.}
Functions whose symbol letters are given by ${\cal S}_{\rm hex,\, MRK}$
in \eqn{eq:ShexMRK}, but with first entries restricted to
$z\zb$ and $(1-z)(1-\zb)$, are called {\it single-valued} HPLs
(SVHPLs)~\cite{BrownSVHPLs,Dixon:2012yy} and denoted by
$\cL_{\vec{w}}(z,\zb)$.  They do not have branch cuts separately
in $z$ and $\zb$; for example, when $z$ is continued around the origin,
and $\zb$ is continued in the opposite direction, the functions remain
single-valued, so they are {\it real analytic} on the punctured complex plane,
$\mathbb{C}/\{0,1,\infty\}$.

The functions $\cL_{\vec{w}}(z,\zb)$ obey same differential relations
in $z$ as $H_{\vec{w}}(z)$:
\beq
\frac{\partial \cL_{0,\vec{w}}(z)}{\partial z}
= \frac{\cL_{\vec{w}}(z)}{z} \,,
\qquad
\frac{\partial \cL_{1,\vec{w}}(z)}{\partial z}
= \frac{\cL_{\vec{w}}(z)}{1-z} \,.
\label{eq:Ldiffdef}
\eeq
They also obey the ``initial conditions''
\beq
\cL(z) = 1, \qquad
\cL_{\vec{0}_n}(z) = \frac{1}{n!} \ln^n |z|^2 \,, \qquad
\lim_{z\rightarrow 0} \cL_{\vec{w} \neq \vec{0}_n}(z) = 0,
\label{eq:L0zdef}
\eeq
and the same shuffle relations as the HPLs,
\begin{equation}
\label{eq:L_shuffle}
\cL_{\vec{w}_1}(z)\,\cL_{\vec{w}_2}(z)
= \sum_{w\in w_1 \ssha w_2} \cL_{\vec{w}}(z)\,,
\end{equation}
These conditions, real analyticity on $\mathbb{C}/\{0,1,\infty\}$,
and the differential relation~(\ref{eq:Ldiffdef}),
fix the $\cL_{\vec{w}}(z,\zb)$.

Each $\cL_{\vec{w}}$ is a weight $|\vec{w}|$ linear combination
of products of HPLs in $z$, HPLs in $\zb$, and multiple zeta values.
The fully holomorphic part of $\cL_{\vec{w}}(z,\zb)$ is
precisely $H_{\vec{w}}(z)$.  The full construction of
$\cL_{\vec{w}}$~\cite{BrownSVHPLs}
relies on the single-valued map ${\bf s}$ mentioned in
sec.~\ref{sec:fmnll}.  It incorporates
the antipode map (which reverses the order of letters in the symbol)
and the Drin'feld associator, which is a formal sum of values
of the HPLs at unit argument, $H_{\vec{w}}(1) = \zeta(\vec{w})$.

It is useful to introduce a {\it collapsed notation} which maps a string of
$(m-1)$ 0's followed by a 1 to the integer $m$:
\beq
\ldots,\vec{0}_{m-1},1,\ldots\ \to\ \ldots,m,\ldots
\label{eq:collapsednotation}
\eeq
In this notation, $\zeta(\vec{w})$ can be identified with the
multiple zeta values (MZVs) defined by the nested sums,
\beq
\zeta_{m_1,\ldots,m_k} = \zeta(m_1,\ldots,m_k) =
\sum_{\infty>i_1>i_2>\cdots>i_k>0} \frac{1}{i_1^{m_1} i_2^{m_2} \cdots i_k^{m_k}} \,.
\label{eq:MZVsumdef}
\eeq

Next we give a few explicit examples of
$\cL_{\vec{w}} \equiv \cL_{\vec{w}}(z,\zb)$~\cite{BrownSVHPLs},
defining $H_{\vec{w}} \equiv H_{\vec{w}}(z)$ and $\Hb_{\vec{w}} = H_{\vec{w}}(\zb)$.
At weight one, there is only
\beq
\cL_{0} \;=\; H_{0} + \Hb_{0} \;=\; \ln |z|^2 \, ,\qquad 
\cL_{1} \;=\;H_{1} + \Hb_{1} \;=\; -\ln|1-z|^2 \,.
\label{eq:Lwt1}
\eeq
The SVHPLs of weight two begin to expose the order-reversal
in the antipode map,
\beqa
\cL_{0,0} &=& H_{0,0} + \Hb_{0,0} + H_{0} \Hb_{0} \,,
\qquad
\cL_{0,1}\;=\;H_{0,1} + \Hb_{1,0} + H_{0} \Hb_{1} \,,
\nn\\
\cL_{1,0} &=& H_{1,0} + \Hb_{0,1} + H_{1} \Hb_{0} \,,
\qquad
\cL_{1,1}\;=\;H_{1,1} + \Hb_{1,1} + H_{1} \Hb_{1} \,.
\label{eq:Lwt2}
\eeqa
Thanks to the shuffle relations~(\ref{eq:HPL_shuffle})
and (\ref{eq:L_shuffle}), it is enough to give results for the
Lyndon basis only, which at weight three is,
\beqa
\cL_{0,0,1}&\;=\;H_{0,0,1}+\Hb_{1,0,0}+H_{0,0}\Hb_{1}+H_{0}\Hb_{1,0}\,,
\nn\\
\cL_{0,1,1}&\;=\;H_{0,1,1}+\Hb_{1,1,0}+H_{0,1}\Hb_{1}+H_{0}\Hb_{1,1}\,,
\label{eq:Lwt3}
\eeqa
and at weight four it is,
\begin{eqnarray}
\hspace{-1cm}
\cL_{0,0,0,1} &=& H_{0,0,0,1}+\Hb_{1,0,0,0}+H_{0,0,0}\Hb_{1}
   +H_{0}\Hb_{1,0,0}+H_{0,0}\Hb_{1,0}\,,
\nn\\
\hspace{-1cm}
\cL_{0,0,1,1} &=& H_{0,0,1,1}+\Hb_{1,1,0,0}+H_{0,0,1}\Hb_{1}
  +H_{0}\Hb_{1,1,0}+H_{0,0}\Hb_{1,1}-2\zeta_3\,\Hb_{1}\,,
\nn\\
\hspace{-1cm}
\cL_{0,1,1,1} &=& H_{0,1,1,1}+\Hb_{1,1,1,0}+H_{0,1,1}\Hb_{1}
  +H_{0}\Hb_{1,1,1}+H_{0,1}\Hb_{1,1}-2\zeta_3\,\Hb_{1}\,,
\label{eq:Lwt4}
\end{eqnarray}
at which point explicit $\zeta$ values from the Drin'feld associator
start to appear.

In summary, the function space for six-point amplitudes in planar
${\cal N}=4$ SYM in MRK reduces from~(\ref{eq:FnsMRK}) to
\beq
{\cal F}_{\rm MRK} \subset 
\left\{ \ln^k(1-u_1),\ k\ge0 \right\} 
\otimes \left\{ \cL_{\vec{w}}(z),\ w_k\in\{0,1\} \right\} \,,
\label{eq:FnsMRKL}
\eeq
including also the possibility of MZV constants.
Note that we can easily trade $(1-u_1)$ for $\sqrt{u_2u_3}$ because the
ratio is finite, from \eqn{eq:zdef},
and its logarithm is in the space of SVHPLs,
\beq
\ln(\sqrt{u_2u_3}) - \ln(1-u_1) = \frac{1}{2} \ln |z|^2 - \ln |1-z|^2
= \frac{1}{2} \cL_0 + \cL_1 \,.
\label{eq:schemeshift}
\eeq
It is convenient to define a large logarithm by
\beq
L_\tau \, \equiv \, \ln\tau, \qquad \tau \equiv \sqrt{u_2u_3} \,.
\label{eq:Ludef}
\eeq

The six-gluon remainder function $R_6(u,v,w)$
vanishes in the Euclidean version of the MRK limit, which is a
soft limit, a special case of the collinear limit, where it also vanishes.
It also vanishes exactly at one loop, by definition.
It is also possible to argue from the OPE perspective
that the leading power of the singular logarithm at $\ell$ loops
should be $\ell-1$~\cite{Alday:2010ku}.
From these properties, and the general structure of the function
space~(\ref{eq:FnsMRKL}),
we can write the general form of the remainder function in MRK
as\footnote{Sometimes the coupling $a = 2g^2$ is used instead of $g^2$.}
\beq
R_6|_{\rm MRK} = 2\pi i \sum_{\ell=2}^\infty \sum_{n=0}^{\ell-1}
g^{2\ell} \ln^n(1-u_1) \Bigl[ g_n^{(\ell)}(z,\zb) + 2\pi i h_n^{(\ell)}(z,\zb) \Bigr] \,,
\label{eq:R6expand}
\eeq
where the imaginary part $g_n^{(\ell)}$ (real part $h_n^{(\ell)}$)
is a weight $2\ell-n-1$ ($2\ell-n-2$) linear combination of SVHPLs.
The overall $2\pi i$ comes from having to take at least a single discontinuity
under the continuation~(\ref{eq:u1continue}); the second discontinuity
contributes to $h_n^{(\ell)}$; and higher odd (even) discontinuities also
contribute to $g_n^{(\ell)}$ ($h_n^{(\ell)}$).  The leading logarithmic (LL)
coefficients clearly have $n=\ell-1$;
the next-to-leading logarithmic (NLL) coefficients have $n=\ell-2$;
and the N$^k$LL coefficients have $n=\ell-k-1$.  Hence the N$^k$LL coefficient
$g_{\ell-k-1}^{(\ell)}$ has weight $\ell+k$.

\begin{figure}
  \centerline{\includegraphics[width=0.32\columnwidth]{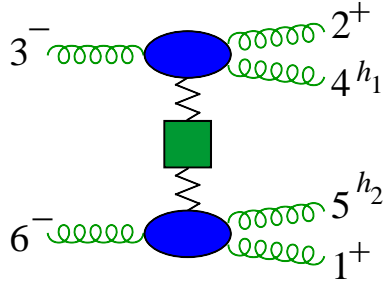} }
  \caption{Factorization of the six-gluon amplitude in planar ${\cal N}=4$
    SYM in the $2\to4$ multi-Regge limit.
    The product of the upper and lower impact factors (blue blobs)
    is given by $\Phi_{\textrm{reg}}$ in the MHV case $(h_1,h_2)=({+},{+})$,
    while the zigzag line and the green square represent the Reggeized gluon
    in the adjoint representation and its BFKL eigenvalue.  In contrast
    to the previous figures, a blob represents all loop orders at once.
    Also, there are vertical cuts through the Reggeized gluon that are not shown.}
\label{fig:Neq4MRK6pt}
\end{figure}

An all-loop integral formula for the six-point amplitude in MRK,
based on factorization in Fourier-Mellin space, was first presented
at NLL in refs.~\cite{Lipatov:2010ad,Fadin:2011we}, and argued to hold
for all subleading logarithms in ref.~\cite{Caron-Huot:2013fea}.
It takes the form (after letting $w=-z$, $\ws=-\zb$)~\cite{DelDuca:2018hrv},
\beqa
\label{eq:MHV_MRK}
\hspace{-1.5cm}
e^{R_6  + i\delta_6}|_{\textrm{MRK}} &=& \cos\Bigl(\pi\Gamma\ln |z|^2 \Bigr) \\
\hspace{-1.5cm}
&&\hskip0cm
+ i \, g^2 \sum_{m=-\infty}^{\infty} \left(\frac{z}{\zb}\right)^{{m\over 2}}
{\cal P} \int_{-\infty}^{\infty}
\frac{d\nu \,|z|^{2i\nu}}{\nu^2+\frac{m^2}{4}}
\Phi_{\textrm{reg}}(\nu,m)
e^{- (L_\tau+i\pi) \omega(\nu,m)} \,,
\nn
\eeqa
where ${\cal P}$ is a principal value prescription for the $m=0$ term,
and $L_\tau$ is defined in \eqn{eq:Ludef}.
The first and second terms are called, respectively, the Regge pole and cut
contributions.
The latter depends on the BFKL eigenvalue $\omega(\nu,m)$ and the (regularized)
impact factor $\Phi_{\textrm{reg}}(\nu,m)$; the latter is really a product
of impact factors for the top and bottom of the Regge ladder shown in
fig.~\ref{fig:Neq4MRK6pt}.
Both $\omega$ and $\Phi_{\textrm{reg}}$ depend on $g^2$.
The function $\delta_6$ appearing on the left-hand side of \eqn{eq:MHV_MRK}
comes from a Mandelstam cut present in the BDS ansatz; it is given by
\beq
\delta_6 = \pi \Gamma(g^2) \, \ln\biggl(\frac{|z|^2}{|1-z|^4}\biggr) \,,
\label{eq:deltadef}
\eeq
where $\Gamma(g^2)$ is defined in \eqn{eq:gamma_cusp}.

In the all-orders solution from the flux-tube
representation~\cite{Basso:2014pla},
the Mellin variable $\nu$ and BFKL eigenvalue $\omega$ are
related to the energy and momenta of an analytically-continued
flux-tube excitation, characterized by a rapidity $u$.
They are both given\footnote{The $\nu$ in \eqn{eq:nufromu} is $1/2$ of
the $\nu$ in ref.~\cite{Basso:2014pla}, in order to be consistent with earlier
definitions of $\nu$.}
in terms of the kernel $K(t)$ entering the BES integral
equation~\cite{Beisert:2006ez},
\beqa
-\omega(u,m) &=& \int_0^{\infty} \frac{dt}{t}
\biggl( \frac{K(-t)+K(t)}{2} \cos(ut) e^{-|m|t/2} - K(t) \biggr) \,,
\label{eq:omegafromu} \\
\nu(u,m) &=& u + \int_0^\infty \frac{dt}{t}
\frac{K(-t)-K(t)}{4} \sin(ut) e^{-|m|t/2} \,.
\label{eq:nufromu}
\eeqa
The impact factor is given in terms of the ``measure'' $\mu^{\rm BFKL}$,
which involves a few more ingredients~\cite{Basso:2014pla}.
The full formula for $R_6$ in MRK is
\beq
\label{eq:MHV_MRK_BCHS}
e^{R_6  + i\delta_6}|_{\textrm{MRK}} =
i \, \sum_{m=-\infty}^{\infty} \left(\frac{z}{\zb}\right)^{{m\over 2}}
\int_{-\infty}^{\infty} du  \, \mu^{\rm BFKL}(u,m)
\, |z|^{2i\nu(u,m)} \, e^{- (L_\tau+i\pi) \omega(u,m)} \,.
\eeq
The Regge pole contribution in \eqn{eq:MHV_MRK} is
not present explicitly in \eqn{eq:MHV_MRK_BCHS}, but it is generated 
by a different contour prescription for the $u$
integral~\cite{Basso:2014pla,DelDuca:2018hrv}.

The BES kernel $K(t)$ can be represented as a semi-infinite matrix
of integrals of products of Bessel functions~\cite{Benna:2006nd}.
At weak coupling, the matrix can be truncated to a finite size,
allowing perturbative expansions of $\nu(u,m)$, $\omega(u,m)$
and $\mu^{\rm BFKL}(u,m)$ to any desired order in the loop
expansion~\cite{Basso:2014pla}.
The results are polynomials in $E,V,N$~\cite{Dixon:2012yy}
defined by
\beqa
E &=& - \frac{1}{2} \frac{|m|}{u^2 + \frac{m^2}{4}}
+ \psi(1+iu+\tfrac{|m|}{2}) + \psi(1-iu+\tfrac{|m|}{2}) - 2 \psi(1),
\label{eq:EVN_Edef} \\
V &=& \frac{iu}{u^2 + \frac{m^2}{4}} \,, \qquad
   N = \frac{m}{u^2 + \frac{m^2}{4}} \,,
\label{eq:EVN_VNdef}
\eeqa
and their $u$ derivatives, with $D \equiv -i\partial/\partial u$.
The results for $\nu$ and $\omega$ through $g^6$ are~\cite{Basso:2014pla},
\beqa
\hspace{-2cm}
2\nu &=& 2u + 2ig^2V - ig^4 (D^2V + 4 \zeta_2 V)
\nn\\
\hspace{-2cm}&&\hskip0.0cm\null
+ ig^6 \biggl( \frac{1}{6} D^4V + 2 \zeta_2 D^2V - 4\zeta_3 DE
+ 44 \zeta_4 V \biggr) + \ldots,
\label{eq:nu3loops}\\
\hspace{-2cm}
-\omega &=& 2g^2E - g^4 ( D^2 E + 4 \zeta_2 E + 12 \zeta_3 )
\nn\\
\hspace{-2cm}&&\hskip0.0cm\null
+ g^6 \biggl( \frac{1}{6} D^4E + 2 \zeta_2 D^2E + 4\zeta_3 DV
+ 44 \zeta_4 E + 80 \zeta_5 + 16 \zeta_2 \zeta_3 \biggr) + \ldots.
\label{eq:E3loops}
\eeqa
Using \eqn{eq:nu3loops},
one can eliminate $u$ in favor of $\nu$ order by order in the coupling,
in order to obtain the more standard definition of the BFKL eigenvalue
$\omega(\nu,m)$ as functions of $E,V,N$ that depend on $\nu$ instead of $u$.
The results agree with previous computations through
NNLL~\cite{Bartels:2008ce,Bartels:2009vkz,Fadin:2011we,
  Dixon:2012yy,Dixon:2014voa}.
The relation between the BFKL measure and the impact factor is
\beq
\mu^{\rm BFKL}(u,m) \, = \, g^2 \frac{d\nu}{du}
\frac{\Phi_{\textrm{reg}}(\nu,m)}{\nu^2 + \frac{m^2}{4}} \,.
\label{eq:muPhimap}
\eeq
This impact factor agrees with previous computations at low loop
orders~\cite{Lipatov:2010ad,Dixon:2012yy,Dixon:2014voa}.

Once the perturbative expansions of $\omega(\nu,m)$ and
$\Phi_{\textrm{reg}}(\nu,m)$ have been obtained, the inverse FM sum-integral
in \eqn{eq:MHV_MRK} has to be performed.  It can be converted into
a double sum by closing the $\nu$ contour with a large semi-circle
in the complex plane and picking up residues from integer-spaced poles on the
positive imaginary axis.
Truncating the sum over residues corresponds to performing a series expansion as
$z,\zb\to0$.  Methods for performing this sum have been given in
refs.~\cite{Dixon:2012yy,Drummond:2015jea,Broedel:2015nfp,DelDuca:2016lad}.
Alternatively, if one has the perturbative amplitude, the coefficient
functions can be computed, without having to perform any sums, by
taking the multi-Regge limit of the result for general kinematics.
It is straightforward to series expand the coefficient functions as
$z,\zb\to0$, to compare with the FM representation.
The results match through seven loops~\cite{Caron-Huot:2019vjl}.

Defining the FM sum $\Sigma$ via,
\beqa
e^{R_6  + i\delta_6}|_{\textrm{MRK}}\ &=&\
\cos\Bigl(\pi\Gamma\ln |z|^2 \Bigr) + \pi i \, \Sigma \,, \nn\\
\hspace{1cm}
\Sigma\ &=&\ \sum_{\ell=1}^\infty g^{2\ell}
           \sum_{n=0}^{\ell-1} \Sigma_{n}^{(\ell)} \, (L_\tau+i\pi)^n \,,
\label{eq:MHV_MRK_Sigma}
\eeqa
the first three loop orders are given in the linear SVHPL representation by,
\beqa
% 1 loop:
\Sigma_{0}^{(1)} &=& \cL_{0}+2 \cL_{1} \,,
\label{FMsum10} \\
% 2 loops:
\Sigma_{1}^{(2)} &=& 2 \cL_{0,1}+2 \cL_{1,0}+4 \cL_{1,1} \,,
\label{FMsum21} \\
\Sigma_{0}^{(2)} &=& -2 \cL_{0,0,1}+2 \cL_{0,1,0}-2 \cL_{0,1,1}
-2 \cL_{1,0,0}-2 \cL_{1,0,1}-2 \cL_{1,1,0}-4 \cL_{1,1,1}
\nn\\ &&\hskip0.0cm\null
-2 \zeta_2 (\cL_{0}+2 \cL_{1}) \,,
\label{FMsum20} \\
% 3 loops:
\Sigma_{2}^{(3)} &=& \cL_{0,0,1}+2 \cL_{0,1,0}+4 \cL_{0,1,1}
+\cL_{1,0,0}+4 \cL_{1,0,1}+4 \cL_{1,1,0}+8 \cL_{1,1,1} \,,
\label{FMsum32} \\
\Sigma_{1}^{(3)} &=& -3 \cL_{0,0,0,1}+\cL_{0,0,1,0}-6 \cL_{0,0,1,1}
+\cL_{0,1,0,0}-4 \cL_{0,1,0,1}-12 \cL_{0,1,1,1}
\nn\\ &&\hskip0.0cm\null
-3 \cL_{1,0,0,0}-8 \cL_{1,0,0,1}-4 \cL_{1,0,1,0}-12 \cL_{1,0,1,1}
-6 \cL_{1,1,0,0}-12 \cL_{1,1,0,1}
\nn\\ &&\hskip0.0cm\null
-12 \cL_{1,1,1,0}-24 \cL_{1,1,1,1}
+4 \zeta_3 \cL_{1}-8 \zeta_2 (\cL_{0,1}+\cL_{1,0}+2 \cL_{1,1}) \,,
\label{FMsum31} \\
\Sigma_{0}^{(3)} &=&
3 \cL_{0,0,0,0,1}-\frac{9}{2} \cL_{0,0,0,1,0}+3 \cL_{0,0,0,1,1}
+5 \cL_{0,0,1,0,0}+3 \cL_{0,0,1,0,1}
\nn\\ &&\hskip0.0cm\null
-\cL_{0,0,1,1,0}+6 \cL_{0,0,1,1,1}+2 \cL_{0,1,0,0,1}
-\frac{9}{2} \cL_{0,1,0,0,0}-2 \cL_{0,1,0,1,0}
\nn\\ &&\hskip0.0cm\null
+4 \cL_{0,1,0,1,1}-\cL_{0,1,1,0,0}+4 \cL_{0,1,1,0,1}
+12 \cL_{0,1,1,1,1}+3 \cL_{1,0,0,0,0}
\nn\\ &&\hskip0.0cm\null
+6 \cL_{1,0,0,0,1}+2 \cL_{1,0,0,1,0}+8 \cL_{1,0,0,1,1}
+3 \cL_{1,0,1,0,0}+8 \cL_{1,0,1,0,1}
\nn\\ &&\hskip0.0cm\null
+4 \cL_{1,0,1,1,0}+12 \cL_{1,0,1,1,1}+3 \cL_{1,1,0,0,0}
+8 \cL_{1,1,0,0,1}+4 \cL_{1,1,0,1,0}
\nn\\ &&\hskip0.0cm\null
+12 \cL_{1,1,0,1,1}+6 \cL_{1,1,1,0,0}+12 \cL_{1,1,1,0,1}
+12 \cL_{1,1,1,1,0}+24 \cL_{1,1,1,1,1}
\nn\\ &&\hskip0.0cm\null
-2 \zeta_2 (3 \cL_{0,0,0}+2 \cL_{0,1,0}-4 \cL_{0,1,1}
           -4 \cL_{1,0,1}-4 \cL_{1,1,0}-8 \cL_{1,1,1})
\nn\\ &&\hskip0.0cm\null
+2 \zeta_3 (\cL_{0,1}+\cL_{1,0}-2 \cL_{1,1})
+22 \zeta_4 (\cL_{0}+2 \cL_{1}) \,.
\label{FMsum30}
\eeqa
Results through seven loops are provided in an ancillary file for
ref.~\cite{Caron-Huot:2019vjl}, although for a slightly different normalization,
coupling-constant convention, and $\cL$ representation.
At LL, only the Regge cut contributes and $e^{R_6} \approx 1+R_6$;
hence the coefficients $\Sigma_{\ell-1}^{(\ell)}$ are simply related to
the corresponding coefficients in the expansion of the remainder function
$R_6$ in \eqn{eq:R6expand}:
\beq
\Sigma_{\ell-1}^{(\ell)} = 2 g_{\ell-1}^{(\ell)} \,.
\label{eq:SigmagRelation}
\eeq
These LL coefficients are known in closed form to all loop
orders~\cite{Pennington:2012zj,Broedel:2015nfp}.

The MRK limit of $3\to3$ scattering is closely related.
There are some sign flips associated with an analytic continuation of $u_1$
in the opposite direction, $u_1 \to u_1 e^{+2\pi i}$,
$u_{2,3} \to u_{2,3} e^{+\pi i}$, and the phase cancels in the exponentiated term,
leading to the following formula~\cite{Bartels:2010tx},
\beqa
\label{eq:MHV_MRK_33}
\hspace{-1.5cm}
e^{R_6  - i\delta_6}|_{\textrm{MRK},\ 3\to3} &=&
\cos\Bigl(\pi\Gamma\ln |z|^2 \Bigr) \\
\hspace{-1.5cm}
&&\hskip0cm
- i \, g^2 \sum_{m=-\infty}^{\infty} \left(\frac{z}{\zb}\right)^{{m\over 2}}
{\cal P} \int_{-\infty}^{\infty}
{d\nu \,|z|^{2i\nu} \over \nu^2+{n^2\over 4}}
\Phi_{\textrm{reg}}(\nu,m)
e^{- L_\tau \omega(\nu,m)} \,.
\nn
\eeqa
Thus the Regge cut term is purely imaginary for $3\to3$ scattering,
and the perturbative results follow from the same FM sum,
\beqa
e^{R_6  - i\delta_6}|_{\textrm{MRK},\ 3\to3} =
\cos\Bigl(\pi\Gamma\ln |z|^2 \Bigr)
- \pi i \sum_{\ell=1}^\infty g^{2\ell}
    \sum_{n=0}^{\ell-1} \Sigma_{n}^{(\ell)} \, (L_\tau)^n \,.
\label{eq:MHV_MRK_Sigma_33}
\eeqa

Flipping the helicity of one of the final-state gluons in
fig.~\ref{fig:Neq4MRK6pt} also results in a minor modification
of the basic formula~(\ref{eq:MHV_MRK}).
Only the impact factor, or BFKL measure, changes.
In terms of the rapidity formulation in
\eqn{eq:MHV_MRK_BCHS}, to flip $h_1$, one simply inserts the factor
\beq
\bar{H}(u,m) \equiv \frac{x(u+\tfrac{im}{2})}{x(u-\tfrac{im}{2})}
\label{eq:MHVtoNMHV}
\eeq
into the $u$ integrand, where
\beq
x(u) = \frac{u+\sqrt{u^2-4g^2}}{2} 
\eeq
is the Zhukovsky variable.  (To flip the helicity $h_2$ of the other more
central gluon, the inverse of the factor~(\ref{eq:MHVtoNMHV}) is used.  The
two cases are related by target-projectile symmetry, which includes
the map $z \to 1/z$, $\zb \to 1/\zb$.)
The NMHV analog of \eqn{eq:MHV_MRK} is then
\beqa
\label{eq:NMHV_MRK}
\hspace{-2.5cm}
{\cal P}^{(4444)}_{\rm NMHV}
e^{R_6  + i\delta_6}|_{\textrm{MRK}} &=& \cos\Bigl(\pi\Gamma\ln |z|^2 \Bigr) \\
\hspace{-2.5cm}
&&\hskip-.2cm
+ i g^2 \! \sum_{m=-\infty}^{\infty} \left(\frac{z}{\zb}\right)^{{m\over 2}}
\! {\cal P} \! \int_{-\infty}^{\infty}
\frac{d\nu \,|z|^{2i\nu}}{\nu^2+\frac{m^2}{4}}
\Phi_{\textrm{reg}}(\nu,m) \bar{H}(u,m)
e^{- (L_\tau+i\pi) \omega(\nu,m)} \,.
\nn
\eeqa
Here ${\cal P}^{(4444)}_{\rm NMHV}$ is the finite {\it ratio function}
of the NMHV super-amplitude divided by the MHV amplitude,
and its $(\eta_4)^4$ Grassmann component according to \eqn{eq:superfield},
in order to flip the helicity of gluon 4,
and $u$ is related to $\nu$ by \eqns{eq:nufromu}{eq:nu3loops}.

While the coefficients in the expansion of the MHV remainder
function~(\ref{eq:R6expand}) are pure transcendental functions,
that is not quite true for the NMHV amplitude.
We define the FM sum $\Sigmat$ via,
\beqa
{\cal P}^{(4444)}_{\rm NMHV} e^{R_6  + i\delta_6}|_{\textrm{MRK}} &=&
\cos\Bigl(\pi\Gamma\ln |z|^2 \Bigr)
 + \pi i \, \frac{\Sigmat(z,\zb) - \zb \, \Sigmat(1/z,1/\zb)}{1-\zb}  \,, \nn\\
\hspace{3cm}
\Sigmat &=& \sum_{\ell=1}^\infty g^{2\ell}
           \sum_{n=0}^{\ell-1} \Sigmat_{n}^{(\ell)} \, (L_\tau+i\pi)^n \,.
\label{eq:NMHV_MRK_Sigmat}
\eeqa
The rational prefactors $1/(1-\zb)$ and $-\zb/(1-\zb)$
arise from the multi-Regge limit of certain dual super-conformal
invariants, or five brackets~\cite{Dixon:2014iba}.
The pure functions in 
$\Sigmat$ can be extracted from the $\zb\to0$ limit of the FM
sum with $\bar{H}$ inserted.  The first few loop orders are,
\beqa
% 1 loop:
\Sigmat_{0}^{(1)} &=& \cL_{0} \,,
\label{FMtsum10} \\
% 2 loops:
\Sigmat_{1}^{(2)} &=& 2 \cL_{0,1} \,,
\label{FMtsum21} \\
\Sigmat_{0}^{(2)} &=& 2 \cL_{0,0,1}-\cL_{0,1,0}-2 \cL_{0,1,1}-2 \zeta_2 \cL_{0} \,,
\label{FMtsum20} \\
% 3 loops: 
\Sigmat_{2}^{(3)} &=& 2 \cL_{0,0,1}+\cL_{0,1,0}+4 \cL_{0,1,1} \,,
\label{FMtsum32} \\
\Sigmat_{1}^{(3)} &=& 
2 \cL_{0,0,1,0}-2 \cL_{0,1,0,0}-4 \cL_{0,1,0,1}-4 \cL_{0,1,1,0}-12 \cL_{0,1,1,1}
-8 \zeta_2 \cL_{0,1} ,
\label{FMtsum31} \\
\Sigmat_{0}^{(3)} &=&
-6 \cL_{0,0,0,0,1}+\frac{9}{2} \cL_{0,0,0,1,0}-3 \cL_{0,0,1,0,0}-2 \cL_{0,0,1,0,1}
-2 \cL_{0,0,1,1,0}
\nn\\ &&\hskip0.0cm\null
+\frac{3}{2} \cL_{0,1,0,0,0}+2 \cL_{0,1,0,0,1}+2 \cL_{0,1,0,1,0}+4 \cL_{0,1,0,1,1}
+2 \cL_{0,1,1,0,0}
\nn\\ &&\hskip0.0cm\null
+4 \cL_{0,1,1,0,1}+4 \cL_{0,1,1,1,0}+12 \cL_{0,1,1,1,1}
\nn\\ &&\hskip0.0cm\null
-2 \zeta_2 (3 \cL_{0,0,0}+2 \cL_{0,0,1}-4 \cL_{0,1,1})
+4 \zeta_3 \cL_{0,1}+22 \zeta_4 \cL_{0} \,.
\label{FMtsum30}
\eeqa
The perturbative expansion has been checked
against bootstrapped NMHV amplitudes through seven
loops~\cite{Dixon:2014iba,Dixon:2015iva,Caron-Huot:2019vjl,
  ToAppearNMHVSevenLoops}.

Multi-Regge kinematics is a particular limit of general $n$-point scattering.
It overlaps with other nearby limits, in particular, those located
at the boundaries of the moduli space of Riemann spheres with marked points.
For $n=6$, there are three such boundaries, for $z \to 0$, 1, or $\infty$.
The $z\to\infty$ limit is a collinear-Regge limit, which is related by
target-projectile symmetry to the $z\to0$ collinear-Regge limit.
The remainder function vanishes in the $z\to0$ limit like $z+\zb$ (or $w+\ws$),
and the leading double logarithms at this power in $\ln|z|^2$ and $L_\tau$
(DLLA) have been summed to all orders
(also for NMHV)~\cite{Bartels:2011xy,Pennington:2012zj}.
The $z\to1$ limit overlaps with a limit in which the dual light-like hexagonal
Wilson loop crosses itself~\cite{Georgiou:2009mp,Dorn:2011gf,Dorn:2011ec},
which is also a limit that mimics double parton scattering,
where a $2\to4$ process breaks up into two $1\to2$ splittings of the incoming
particles, followed by two $2\to2$ scatterings~\cite{Dixon:2016epj}.
The singular terms in this limit can be understood to all
orders, and also in the related $3\to3$ version of
self-crossing~\cite{Dixon:2016epj,Caron-Huot:2019vjl}.

\subsection{Seven-gluon MRK}
\label{sec:SevengluonMRK}

\begin{figure}
  \centerline{\includegraphics[width=0.32\columnwidth]{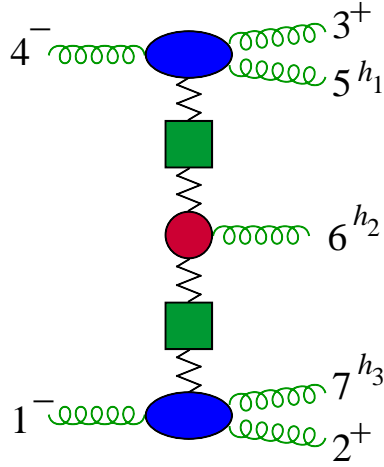} }
  \caption{Factorization of the seven-gluon amplitude in planar ${\cal N}=4$
    SYM in the $2\to5$ multi-Regge limit.
    The additional factor not present in the six-point case is
    the central emission vertex (red blob), which again represents all orders.
    Again the vertical Regge cuts are omitted.}
\label{fig:Neq4MRK7pt}
\end{figure}

The multi-Regge limit of $2\to5$ scattering features five particles
with strongly-ordered rapidities in the final state.  The configuration
with the richest dynamics is where the particles with maximal separation
in the cyclic color ordering are incoming, say particles 1 and 4,
and particles 2, 3, 5, 6, and 7 are outgoing,
as shown in fig.~\ref{fig:Neq4MRK7pt}.  Alternatively, we could start
with particles 2 and 3 incoming, and strongly order the outgoing rapidities,
\beq
p_4^+ \gg p_5^+ \gg p_6^+ \gg p_7^+ \gg p_1^+ \,,
\label{eq:sevenfinalordered}
\eeq
with comparable transverse momenta,
\beq
|p_{4\perp}| \simeq |p_{5\perp}| \simeq |p_{6\perp}|
\simeq |p_{7\perp}|  \simeq |p_{1\perp}| \,,
\label{eq:mrk7pt}
\eeq
and then analytically continue particles 5, 6, 7 to negative energies.
This region features a long Regge
cut~\cite{Bartels:2011ge,Bartels:2013jna,Bartels:2014jya}.

Dual conformal invariance implies that there are only $3\times 7 - 15 = 6$
independent variables for the seven-point remainder function.
There are seven different dual conformal cross ratios, 
\beqa
u_1 &=& \frac{s_{34}s_{671}}{s_{234}s_{345}} \,, \qquad
u_2 = \frac{s_{45}s_{712}}{s_{345}s_{456}} \,, \qquad
u_3 = \frac{s_{56}s_{123}}{s_{456}s_{567}} \,, \qquad
u_4 = \frac{s_{67}s_{234}}{s_{567}s_{671}} \,, \nonumber\\[7pt]
u_5 &=& \frac{s_{71}s_{345}}{s_{671}s_{712}} \,, \qquad
u_6 = \frac{s_{12}s_{456}}{s_{712}s_{123}} \,, \qquad
u_7 = \frac{s_{23}s_{567}}{s_{123}s_{234}} \,,
\label{eq:uisi}
\eeqa
with one nonlinear Gram determinant relation between them.
In MRK, they behave as
\beq
u_1\,,u_2\,,u_5\,,u_6 \sim \cO(\delta), \qquad
1-u_3\,,1-u_4 \sim \cO(\delta), \qquad
1-u_7 \sim \cO(\delta^2) \,,
\label{eq:MRKscaling}
\eeq
where $\delta\to0$, and the small ratios $p_{i+1}^+/p_i^+$ in
\eqn{eq:sevenfinalordered} are all $\cO(\delta)$, for $i=4,5,6,7$.

This limit, which is to be taken after the analytical continuation
\beq
u_7 \to u_7 e^{-2\pi i} \,,
\label{eq:u7continue}
\eeq
can be parametrized by two small real parameters $\tau_{1,2}$
and two complex variables $z_{1,2}$:
\beqa
\label{eq:MRKvariables}
\sqrt{u_1 u_2} &=& \tau_1 \,,  \qquad
\frac{u_1}{1-u_3} = \left|\frac{1}{1-z_1}\right|^2, \qquad
\frac{u_2}{1-u_3} = \left|\frac{z_1}{1-z_1}\right|^2,
\nonumber \\
\sqrt{u_5 u_6} &=& \tau_2 \,,  \qquad
\frac{u_5}{1-u_4} = \left|\frac{1}{1-z_2}\right|^2, \qquad
\frac{u_6}{1-u_4} = \left|\frac{z_2}{1-z_2}\right|^2 .
\eeqa
The \emph{simplicial coordinates} $\rho_1,\rho_2$ are also
used~\cite{Brown:2009qja,DelDuca:2016lad};
they are related to $z_1,z_2$ by
\beq
\label{eq:simplicial}
z_1 = \frac{\rho_1 (1-\rho_2)}{\rho_1 - \rho_2}  \,, \qquad
z_2 = \frac{\rho_2 - \rho_1}{1 - \rho_1} \,.
\eeq

There are 42 letters in the heptagon function symbol
alphabet~\cite{Golden:2013xva,Drummond:2014ffa}.
In MRK they collapse to
\beq
{\cal S}_{\rm hept,MRK} = \{ \tau_{1,2} \,,
    \rho_{1,2} \,, 1-\rho_{1,2} \,, \rho_1-\rho_2 \,,
    \bar\rho_{1,2} \,, 1-\bar\rho_{1,2} \,, \bar\rho_1-\bar\rho_2 \} \,.
\label{eq:SheptMRK}
\eeq
In analogy to the six-point case, the parity-even letters only
involve $\tau_{1,2}$ and the five magnitudes,
\beq
\{ |\rho_{1,2}|^2 \,, |1-\rho_{1,2}|^2 \,, |\rho_1-\rho_2|^2 \} \,.
\label{eq:SheptMRKeven}
\eeq
Only these parity-even combinations appear at the front of symbols, even after
clipping off multiple $u_7$ initial entries~\cite{Dixon:2021nzr},
in accordance with the continuation~(\ref{eq:u7continue}).
Therefore the relevant function space is SVHPLs in two variables, $\rho_1$
and $\rho_2$~\cite{Broedel:2016kls,DelDuca:2016lad}.
(In ref.~\cite{DelDuca:2016lad} it has been argued that
single-valued functions in multiple variables
are all that is needed for MRK for any $n$-point amplitudes, at least at LL.)

An all-orders formula for the multi-Regge limit of any $n$-point amplitude
in planar ${\cal N}=4$ SYM has been proposed based on some
of the same ingredients encountered at $n=6$, plus a new ingredient,
the central emission vertex~\cite{DelDuca:2019tur}.
The formula for $n=7$ is
\beqa
\hspace{-2.55cm}
\cR_{h_1,h_2,h_3} \, e^{i\delta_7} &=& 1 + 2\pi i \, \prod_{k=1,2}
\left[
\sum_{m_k=-\infty}^\infty \left(\frac{z_k}{\bar{z}_k}\right)^{\frac{m_k}{2}}
\int_{\cC_k} \frac{d\nu_k}{2\pi} |z_k|^{2i\nu_k} \tilde{\Phi}(\nu_k,m_k)
    e^{-(\ln\tau_k + i\pi)\omega(\nu_k,m_k)}
    \right]
\nn \\
\hspace{-2.55cm}
&&\null \times
\bigg[
  I^{h_1}(\nu_1,m_1) \cdot \tilde{C}^{h_2}(\nu_1,m_1,\nu_2,m_2)
  \cdot \bar{I}^{h_3}(\nu_2,m_2)
\bigg] \,.
\label{eq:FMintegral7}
\eeqa
Here $\cR_{h_1,h_2,h_3}$ is the appropriate helicity amplitude, divided
by $A_7^{\rm BDS}$. The seven-point BDS phase is
\beq \label{eq:delta7}
\delta_7\ =\ \pi \Gamma \ln \left|\frac{\rho_1}{(1-\rho_1)(1-\rho_2)}\right|^2 \,.
\eeq
The quantity
$\tilde{\Phi}(\nu,m) \equiv g^2 \Phi_{\textrm{reg}}(\nu,m)/(\nu^2+m^2/4)$;
$I^h$ is the helicity flip kernel (related to $\bar{H}$);
and $\tilde{C}^{h}$ is the central emission vertex,
whose all-orders expression is given in ref.~\cite{DelDuca:2018hrv},
along with details about performing the $\nu_k$ integration.
Note that $\tilde{\Phi}$ ($\tilde{C}^{h}$)
does not correspond precisely to the blue (red) blob in fig.~\ref{fig:Neq4MRK7pt},
since there are $n-5$ $\tilde{\Phi}_r$'s and only two true impact factors
for any $n$.
However, it is possible to associate a ``square-root'' of each $\tilde{\Phi}_r$
with a neighboring $\tilde{C}^{h}$ if one wants to make the correspondence
more exact.

As mentioned in the introduction, seven-point amplitudes have
been bootstrapped through four loops at the symbol
level~\cite{Drummond:2014ffa,Dixon:2016nkn,Drummond:2018caf},
and more recently at the level of full functions~\cite{Dixon:2020cnr},
by fixing zeta-valued constants of integration in the Euclidean region.
The proposal~(\ref{eq:FMintegral7}) was checked first at symbol
level~\cite{DelDuca:2019tur}, and more recently at function level,
for both MHV and NMHV configurations, by carrying the constants of integration
along a multi-step path from the Euclidean region to the $2\to5$ multi-Regge
limit~\cite{Dixon:2021nzr}.

%%%%%%%%%%%%%%%%%%%%%%%%%%%%%%%%%%%%%%%%%%%%%%%%%%%%%%%%%%%%%%%%%%%

\section{Conclusions and Outlook}
\label{sec:concl}

In this chapter, we have reviewed the behavior of amplitudes in QCD and in ${\cal N}=4$ SYM in the multi-Regge limit.
In sec.~\ref{sec:qcdregge}, we have detailed how at LL and NLL accuracy QCD amplitudes are dominated by single-Reggeized-gluon exchange.
At NNLL accuracy, three-Reggeized-gluon exchange occurs in the $2\to 2$ amplitude, starting at two loops. In the two-loop five-point amplitude,
the central-emission vertex may couple to the three Reggeized gluons. It would be interesting to understand how that comes about, perhaps first
in the context of full-color ${\cal N}=4$ SYM~\cite{Caron-Huot:2020vlo}. The three-Reggeized-gluon exchange yields a violation of Regge factorization
only at the level of terms that are subleading in $N_c$. Basically, multi-Reggeon contributions seem to be washed away by the large $N_c$ limit. It is conceivable then that radiative corrections might be resummable in QCD through a BFKL equation at NNLL, at least in the large $N_c$ limit. Indeed, in planar ${\cal N}=4$ SYM the color-singlet BFKL eigenvalue has recently been obtained through NNNLL using quantum spectral curve methods~\cite{Alfimov:2018cms,Velizhanin:2021bdh}.

In sec.~\ref{sec:planarN4regge}, we have analyzed the multi-Regge limit of amplitudes with six or more points in planar ${\cal N}=4$ SYM, 
in the long Regge cut configuration, characterized by an energy-sign flip of all the gluons emitted along the gluon ladder, which amounts to
all the produced gluons except the first and the last in the strong rapidity ordering. The energy configuration of the outgoing gluons is then $(+,-,-,+)$ at six points, $(+,-,-,-,+)$ at seven points, and so forth.\footnote{These energy configurations are distinct from helicity configurations, of course.}
In this configuration, the amplitudes feature a two-Reggeized-gluon exchange~\cite{Bartels:2008ce,Bartels:2009vkz}, 
and are conjecturally known for any leg multiplicity and at any logarithmic accuracy~\cite{DelDuca:2019tur}.

In the integrability picture of ${\cal N}=4$ SYM in the large $N_c$ limit in MRK~\cite{Lipatov:2009nt},
amplitudes which feature the exchange of an $n$-Reggeized-gluon ladder are related to an $n$-site Hamiltonian of an integrable
spin chain. The amplitudes with exchange of a two-Reggeized-gluon ladder have been described in sec.~\ref{sec:planarN4regge}.
The two-site Hamiltonian is the BFKL Hamiltonian~\cite{Lipatov:1993qn}. Therefore, those aspects of the integrability picture are non-trivially probed
with the exchange of three or more Reggeized gluons.
According to the integrability picture, amplitudes with eight or more points feature the exchange of a three-Reggeized-gluon ladder in the energy 
configuration characterized by a further energy-sign flip of the innermost gluons emitted along the gluon ladder~\cite{Bartels:2011nz}.
The energy configuration of the outgoing gluons is then $(+,-,+,+,-,+)$ at eight points, $(+,-,+,+,+,-,+)$ at nine points, and so on.
These contributions should have double discontinuities in the same channel, associated with pairs of vertical cuts.
The exchange of a four-Reggeized-gluon ladder
(associated with triple discontinuities)
would appear in amplitudes with ten or more points featuring a further energy-sign flip of the innermost gluons, such that the energy configuration would be $(+,-,+,-,-,+,-,+)$ at ten points, $(+,-,+,-,-,-,+,-,+)$ at eleven points. One can continue flipping energies and considering more Reggeized gluons being exchanged.

Amplitudes at the two-loop level and up to nine points have been analyzed by lifting the symbol of ref.~\cite{Caron-Huot:2011zgw} to function level in the multi-Regge limit. For the regions $(+,-,+,+,-,+)$ and $(+,-,+,+,+,-,+)$ at eight and nine points, respectively, results are consistent with the picture of an amplitude made from impact factors and a central-emission vertex involving the exchange of a three-Reggeized-gluon ladder~\cite{DelDuca:2018raq}.
That is a good start for Lipatov's integrability picture~\cite{Lipatov:2009nt}. Much more, though, remains to be understood.

There are many other avenues for future research on the multi-Regge limit of gauge theory.  They include further developments at NNLL in QCD, including also computations of impact factors and developments in the phenomenological application of these results (which we have not been able to review here).  Beyond the planar limit, ${\cal N}=4$ SYM may provide a useful testing ground for untangling multi-Reggeized-gluon ladders from each other.  Within the planar limit, MRK can serve as a window into the complexity of multi-loop $n$-point amplitudes for generic kinematics: the elliptic polylogarithms and more complicated functions that are expected to be encountered there must simplify drastically in MRK. Exactly how this works remains to be understood.
All in all, the study of the multi-Regge limit in gauge theory will undoubtedly continue to be a rich mother lode within the field of scattering amplitudes.

%%%%%%%%%%%%%%%%%%%%%%%%%%%%%%%%%%%%%%%%%%%%%%%%%%%%%%%%%%%%%%%%%%%

\section*{Acknowledgments}

We thank Andy Liu for a careful reading of the manuscript.
This work was supported by the European Union’s Horizon 2020 research
and innovation programme under the Marie Sk\l{}odowska-Curie grant
agreement No.~764850 {\it ``\href{https://sagex.org}{SAGEX}''},
and by the US Department of Energy under contract DE--AC02--76SF00515.
The figures were drawn with {\sc JaxoDraw}~\cite{Binosi:2008ig}.\\

% \bibliography{ljdvdd.bib}

\providecommand{\newblock}{}

\end{document}